\documentclass[reprint,aps,pre,superscriptaddress]{revtex4-1}

\usepackage{graphicx}
\usepackage{bm}
\usepackage{hyperref}
\usepackage[latin1]{inputenc}
\usepackage{amsmath}
\usepackage{amsfonts}
\usepackage{amssymb}
\usepackage{amsthm}

\makeatletter
\newcommand{\colorcaption}[2][]{%
  \begingroup%
  \renewcommand{\@caption@fignum@sep}{ (Color online). }%
  \caption[#1]{#2}%
  \endgroup%
}
\makeatother

\begin{document}


\title{Laminar Chaos in experiments and nonlinear delayed Langevin equations: A time series analysis toolbox for the detection of Laminar Chaos}

\author{David M\"uller-Bender}
 \email{david.mueller@physik.tu-chemnitz.de}
 \affiliation{Institute of Physics, Chemnitz University of Technology, 09107 Chemnitz, Germany}
\author{Andreas Otto}%
 \email{otto.a@mail.de}
 \affiliation{Institute of Physics, Chemnitz University of Technology, 09107 Chemnitz, Germany}
\author{G\"unter Radons}
 \email{radons@physik.tu-chemnitz.de}
 \affiliation{Institute of Physics, Chemnitz University of Technology, 09107 Chemnitz, Germany}
\author{Joseph D. Hart}
 \email{jdhart12@gmail.com}
 \affiliation{Institute for Research in Electronics and Applied Physics, University of Maryland, College Park, Maryland 20742, USA}
 \affiliation{Department of Physics, University of Maryland, College Park, Maryland 20742, USA}
\author{Rajarshi Roy}%
 \email{rroy@umd.edu}
 \affiliation{Institute for Research in Electronics and Applied Physics, University of Maryland, College Park, Maryland 20742, USA}
 \affiliation{Department of Physics, University of Maryland, College Park, Maryland 20742, USA}
 \affiliation{Institute for Physical Science and Technology, University of Maryland, College Park, Maryland 20742, USA}
%


\date{\today}

\begin{abstract}
Recently, it was shown that certain systems with large time-varying delay exhibit different types of chaos, which are related to two types of time-varying delay: conservative and dissipative delays.
The known high-dimensional Turbulent Chaos is characterized by strong fluctuations.
In contrast, the recently discovered low-dimensional Laminar Chaos is characterized by nearly constant laminar phases with periodic durations and a chaotic variation of the intensity from phase to phase.
In this paper we extend our results from our preceding publication [J. D. Hart, R. Roy, D. M\"uller-Bender, A. Otto, and G. Radons, PRL 123 154101 (2019)], where it is demonstrated that Laminar Chaos is a robust phenomenon, which can be observed in experimental systems.
We provide a time series analysis toolbox for the detection of robust features of Laminar Chaos.
We benchmark our toolbox by experimental time series and time series of a model system which is described by a nonlinear Langevin equation with time-varying delay.
The benchmark is done for different noise strengths for both the experimental system and the model system, where Laminar Chaos can be detected, even if it is hard to distinguish from Turbulent Chaos by a visual analysis of the trajectory.
\end{abstract}

\maketitle

\section{Introduction}
\label{sec:intro}

In nature or in physical experiments noise is always present and its influence has to be taken into account for a realistic description of these processes.
If a theoretically proposed phenomenon is not robust against noise, in the sense that certain characteristics can not be measured in the presence of noise, it is unlikely to observe this phenomenon in nature or to verify the phenomenon experimentally.
Systems that involve transport processes over finite distances by finite velocities are characterized by time-delays.
This is the case, for example, in engineering processes \cite{insperger_semi-discretization_2011} such as turning \cite{otto_application_2013,otto_influence_2015} and milling \cite{otto_stability_2017}, life sciences \cite{smith_introduction_2010}, and climate science \cite{tziperman_nino_1994,ghil_delay_2008,quinn_mid-pleistocene_2018}.
Time-delays are also present in population dynamics where the individual organisms need to reach a maturity threshold in order to get the ability to reproduce \cite{gopalsamy_stability_1992}.
The high dimensional nature and the large variety of dynamical behaviors of systems with delays is exploited in several applications such as the efficient implementation of reservoir computing via time-delay systems \cite{appeltant_information_2011,brunner_parallel_2013,larger_high-speed_2017,brunner_tutorial:_2018}.
An overview of recent developments in the field of nonlinear time-delay systems can be found in the theme issue introduced in Ref.~\cite{otto_nonlinear_2019} and an extensive review of chaos in systems with time-delay is given in Ref.~\cite{wernecke_chaos_2019}.
Due to the importance of these concepts, it seems consequent that the influence of noise on systems with time-delay has also drawn attention in the literature, where analyses of stochastic linear \cite{kuchler_langevins_1992,budini_functional_2004,mcketterick_exact_2014,giuggioli_fokkerplanck_2016} and nonlinear systems \cite{guillouzic_rate_2000,frank_analytical_2004,frank_delay_2005,frank_delay_2005-1,frank_time-dependent_2006,brett_stochastic_2013,caceres_passage_2014,rosinberg_stochastic_2015} with delay can be found.

Due to environmental fluctuations, time-delays are typically not constant, but rather time-varying.
If a system involves transport processes with time-varying velocity or time-varying transport distance, the delay is time-varying \cite{otto_transformations_2017,muller_resonant_2019}.
Although models with time-varying delay are often more realistic, there are only a few papers on such systems with and without noise.
A fast time-varying delay can be approximated by a distributed delay as introduced in Ref.~\cite{michiels_stabilization_2005} and applied in Ref.~\cite{michiels_stabilization_2005,gjurchinovski_stabilization_2008,gjurchinovski_variable-delay_2010, jungling_experimental_2012,gjurchinovski_delayed_2013,gjurchinovski_amplitude_2014, sugitani_delay-_2015}.
It is known that a time-varying delay can enrich the dynamics of systems compared to systems with constant delay \cite{radons_complex_2009,martinez-llinas_dynamical_2015, lazarus_dynamics_2016, grigorieva_chaotic_2018,otto_delay-induced_2017}.
A time-varying delay has an influence on chaos and synchronization, which is discussed in Ref.~\cite{kye_synchronization_2004, kye_characteristics_2004, kye_synchronization_2004_2, ghosh_synchronization_2007} together with possible applications to chaos communication.
For applications in control theory, it is of interest that a time-varying delay can stabilize systems \cite{madruga_effect_2001,insperger_stability_2004, otto_stability_2011, otto_application_2013}.
The stability of systems with stochastically varying delay is analysed in Ref.~\cite{verriest_stability_2009}.

In this paper and in our preceding publication Ref.~\cite{hart_laminar_2019}, we extend our results in Ref.~\cite{muller_laminar_2018,muller_resonant_2019}, where we have introduced a new type of chaos, which can be observed only in systems with certain time-varying delays.
This type of chaos is called \emph{Laminar Chaos} and is characterized by nearly constant laminar phases of periodic duration, where the intensity levels of the laminar phases vary chaotically.
It may be possible to exploit the sequence of nearly constant intensity levels to store information or perform computations, where approaches similar to the chaos based logic and computation introduced in Ref.~\cite{sinha_dynamics_1998,murali_logic_2009,ditto_chaogates:_2010,ditto_exploiting_2015,del_hougne_leveraging_2018} may be developed.
However, this and similar applications require that the sequence of intensity levels is robust against noise, which is hitherto unclear.
We demonstrate that Laminar Chaos is a robust phenomenon, which can be observed experimentally and survives the presence of noise.
In order to do this we have implemented an opto-electronic system with time-varying delay that shows Laminar Chaos and substantiate our results by modelling the system by a nonlinear delayed Langevin equation with time-varying delay.
By the experimental and theoretical analysis we justify that Laminar Chaos has characteristic properties that survive for even comparably large noise strengths and we provide a toolbox for the detection of these features in experimental time series, where the underlying system needs not to be known.

In Sec.~\ref{sec:reviewlamchaos} we give a short review of the results in Ref.~\cite{muller_laminar_2018}.
We introduce the concept of conservative and dissipative delays, which are classes of time-varying delays that lead to certain dynamical properties of the involved systems, and explain Laminar Chaos.
In Sec.~\ref{sec:exprealize} we explain the experimental realization of Laminar Chaos by an opto-electronic setup similar to the one in Ref.~\cite{hart_experiments_2017}.
Results of the experiments are presented, where we demonstrate that Laminar Chaos is robust against experimental and measurement noise.
In Sec.~\ref{sec:model} we derive a general model for our experimental system, which is the basis of the following theoretical analysis of Laminar Chaos in the presence of noise.
A theoretical analysis of the robust features of Laminar Chaos in the presence of noise is done in Sec.~\ref{sec:lamchaosnoise}.
There we show how Laminar Chaos can be identified in experimental time series and present robust features, which can be detected even in the presence of strong noise, where the time series can not be identified as Laminar Chaos visually.

\section{Review of Laminar Chaos}
\label{sec:reviewlamchaos}

In Ref.~\cite{muller_laminar_2018} a new type of chaos, called Laminar Chaos, was found in systems described by the delay differential equation (DDE)
\begin{equation}
\label{eq:sysdef}
\frac{1}{T} \dot{z}(t) + z(t) = \mu \, f(z(R(t))),
\end{equation}
where $R(t)=t-\tau(t)$ is the retarded argument and $\tau(t)$ is a periodically time-varying delay with period one.
In the following we give a short review of the theory behind the dynamics of systems with time-varying delay and Laminar Chaos.
In Ref.~\cite{otto_universal_2017,muller_dynamical_2017,muller_dynamical_2018} it was demonstrated for invertible $R(t)$, i.e., we have $\dot{\tau}(t)<1$ for almost all $t$, that there are two classes of periodically time-varying delays.
We call them \emph{conservative delays} and \emph{dissipative delays}, where the terms, respectively, refer to marginally stable quasiperiodic dynamics or stable periodic dynamics of the access map
\begin{equation}
\label{eq:accessmap}
t=R(t')
\end{equation}
taken modulo the period of the delay.
Systems with a so-called conservative delay are equivalent to systems with constant delay.
This means that there exists a suitable nonlinear timescale transformation, which transforms Eq.~\eqref{eq:sysdef} to a DDE with constant delay and preserves dynamical quantities such as Lyapunov exponents.
In contrast, systems with dissipative delay are not equivalent to constant delay systems and the characteristics of the dynamics differ from the characteristics known for systems with constant delay.
For instance, there are qualitative differences in the asymptotic scaling behavior of the Lyapunov spectrum \cite{otto_universal_2017,muller_dynamical_2017,muller_dynamical_2018} and in the localization properties of the Lyapunov vectors \cite{muller_dynamical_2017}.
Beside Laminar Chaos, systems with dissipative delay show further types of dynamics that are not possible for systems with constant delay such as generalized Laminar Chaos and time-multiplexed dynamics as demonstrated in Ref.~\cite{muller_resonant_2019}.
If the time-varying delay of a system is caused by a transport process over a constant distance by a variable velocity, the time-varying delay is always conservative \cite{otto_transformations_2017}.
On the other hand, if the distance is also time-varying, both classes, conservative and dissipative delays can be observed \cite{muller_resonant_2019}.

From this point we follow the analysis in Ref.~\cite{muller_laminar_2018} and consider Eq.~\eqref{eq:sysdef} with large $T$.
A large parameter $T$ is equivalent to a large time-varying delay, which can be shown easily by a linear time-scale transformation as done in Sec.~\ref{sec:model}.
For $T\to\infty$, the left hand side of Eq.~\eqref{eq:sysdef} vanishes and we obtain the limit map
\begin{equation}
\label{eq:limitmap}
z_{n+1}(t)=\mu \, f(z_n(R(t))).
\end{equation}
The $z_n(t)=z(t)$ with $t\in \mathcal{I}_n=(t_{n-1},t_n]$ are segments of the solution $z(t)$, where the boundaries $t_n$ of the so-called state-intervals $\mathcal{I}_n$ are determined by the access map Eq.~\eqref{eq:accessmap}, i.e., $t_{n-1}=R(t_n)$.
With a given $z_0(t), \; t\in \mathcal{I}_0$ the solution $z(t)$ can be generated by successive iteration of Eq.~\eqref{eq:limitmap}.
This is similar to the method of steps \cite{bellman_computational_1965}, which can be applied in the case $T<\infty$.
As in the case of a constant delay \cite{ikeda_high-dimensional_1987}, for $T<\infty$, one step of the method of steps can be decomposed into one iteration of the limit map and subsequent smoothing or low pass filtering.
This can be seen directly in Eq.~\eqref{eq:sysdef}, since the left hand side defines a low pass filter with cutoff frequency $f_{\text{cutoff}}=T/(2\pi)$ and the remaining terms determine the limit map.
Eq.~\eqref{eq:limitmap} can be also interpreted as the iteration of the graph $(t,z_n(t))$ by the two-dimensional map
\begin{subequations}
\begin{align}
t' &= R^{-1}(t), \label{eqn:2dmapb}\\
z' &= \mu \, f(z), \label{eqn:2dmapa}
\end{align}
\end{subequations}
which consists of two one-dimensional maps:
The map $z'=\mu \, f(z)$ and the inverse of the access map, given by Eq.~\eqref{eq:accessmap}.
So it is clear that, in particular for large $T$, these maps play a crucial role for the dynamics of the system defined by DDE~\eqref{eq:sysdef}.
In the following we consider a chaotic map $z'=\mu \, f(z)$.

For conservative delays $\tau(t)$ the access map taken modulo the period of the delay and its inverse, which is given by Eq.~\eqref{eqn:2dmapb} taken modulo the period of the delay, is topological conjugate to a pure rotation \cite{otto_universal_2017}, i.e., it is a conservative system that shows quasiperiodic dynamics.
This means, roughly speaking, that the map defined by Eq.~\eqref{eqn:2dmapb} and the time-varying delay $\tau(t)$ has no influence on the dynamics of the limit map except for a quasiperiodic frequency modulation \cite{muller_resonant_2019}.
Due to the stretching and folding by the map $z'= F(z) = \mu \, f(z)$, the limit map dynamics is characterized by oscillations, whereby the characteristic frequency of a solution segment $z_n(t)$ grows with $n$, i.e, it grows with each iteration of the limit map.
For finite $T$ the frequency is bounded by the smoothing, i.e., by the low pass filter, leading to a type of chaos that is already known for systems with constant delay.
We call this type of chaos Turbulent Chaos based on the term Optical Turbulence, which was introduced in Ref.~\cite{ikeda_optical_1980}.
An exemplary trajectory of DDE~\eqref{eq:sysdef} that shows Turbulent Chaos is plotted in Fig.~\ref{fig:traj}d.
Turbulent chaos is characterized by a large attractor dimension, since the dimension grows typically linearly with $T$ \cite{farmer_chaotic_1982,ikeda_high-dimensional_1987}.

\begin{figure}[!ht]
\includegraphics[width=0.4\textwidth]{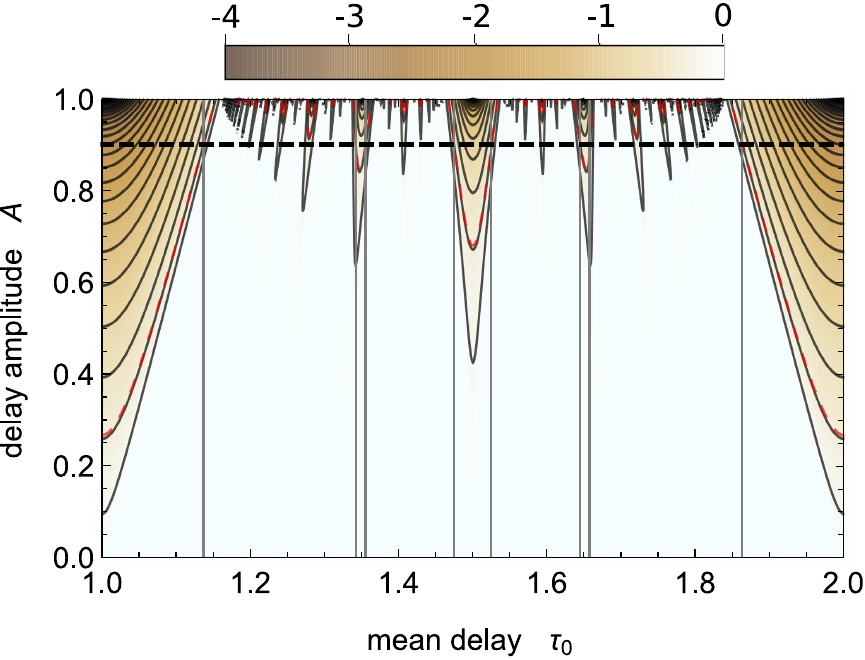}
\caption{\label{fig:crit}
Visualization of the criterion~\eqref{eq:crit} for Laminar Chaos by a heat map of the Lyapunov exponent of the access map, which corresponds to the sinusoidal delay $\tau(t)= \tau_0 + A \sin(2\pi\,t)/(2\pi)$.
The contour lines correspond to a constant Lyapunov exponent (from bottom to top $\lambda[R]=-0.1,-0.3,-0.5,\dots$).
For Eq.~\eqref{eq:sysdef} with $\mu=2.2$, $f(z)= \sin^2(z+\phi_0)$, and $\phi_0=\frac{\pi}{4}$ the criterion~\eqref{eq:crit} is fulfilled above the dashed red line, which corresponds to the contour line $\lambda[R]=-\lambda[F]\approx-0.31$.
The vertical lines represent regions of $\tau_0$ where the criterion~\eqref{eq:crit} is fulfilled for a fixed delay amplitude $A=0.9$.
}
\end{figure}

For dissipative delays the access map modulo the period of the delay has stable fixed points or periodic orbits, i.e., it is a dissipative system.
Points that are evolved by its inverse Eq.~\eqref{eqn:2dmapb} accumulate at the repulsive periodic points $t^*_k$ ($k\in\mathbb{Z}$) of the access map, which are solutions of $R^q(t^*_k)+p=t^*_k$, with $(R^q)'(t^*_k)>0$, where $\rho=-\frac{p}{q}$ is the rotation number (cf. Ref.~\cite{katok_introduction_1997}) of Eq.~\eqref{eqn:2dmapb}.
It follows that points $(t_n,z_n)$ that are evolved under the iterations of the two-dimensional map defined by Eq.~\eqref{eqn:2dmapb} and \eqref{eqn:2dmapa} accumulate at vertical lines with $t=t^*_{k(n)}$, i.e., $(t_n,z_n)\to (t^*_{k(n)},z_n)$.
Consequently, the strong oscillations that are caused by the map $z'=\mu \, f(z)$ during the iteration of the limit map accumulate at these points.
For finite $T$, these oscillations are smoothed and, in the case of Laminar Chaos, they form the possibly irregular transitions between the laminar phases.
A trajectory that shows Laminar Chaos is visualized in Fig.~\ref{fig:traj}a. 
In Ref.~\cite{muller_laminar_2018} it was shown that a necessary condition for Laminar Chaos can be derived from the derivative of the limit map and is given by
\begin{equation}
\label{eq:crit}
\lambda[F]+\lambda[R]<0,
\end{equation}
where $\lambda[F]$ and $\lambda[R]$ are the Lyapunov exponents of the map $z'=\mu\,f(z)$ and the access map $t'=R(t)$, respectively.
This criterion is sufficient for $T=\infty$ and is visualized in Fig.~\ref{fig:crit}.
If Eq.~\eqref{eq:crit} is fulfilled, the oscillations caused by the map $z'=\mu\,f(z)$ are compensated by the contraction of the access map, and the laminar phases can develop between the repulsive periodic points $t^*_k$ of the access map.
Since the repulsive periodic points $t^*_k$ separate $p$ basins of attraction of periodic points inside each state interval $\mathcal{I}_n$, each of the latter contains $p$ laminar phases (see Ref.~\cite{muller_laminar_2018,muller_resonant_2019} for details).
Since the access map is invertible, the intensity levels of the laminar phases inside the state interval $\mathcal{I}_n$ are successively mapped to the intensity levels in $\mathcal{I}_{n+1}$ by $z'=\mu\,f(z)$.
Laminar Chaos is characterized by comparably low attractor dimension, since the state $z_n(t)$ is merely determined by the intensity levels of the laminar phases.
If Eq.~\eqref{eq:crit} is not fulfilled, one can observe generalized Laminar Chaos, which is also a low-dimensional phenomenon compared to Turbulent Chaos \cite{muller_resonant_2019}.

\section{Experimental realization of variable delay systems and Laminar Chaos}
\label{sec:exprealize}

In this section, we describe how Eq.~\eqref{eq:sysdef} with time-varying delay can be implemented in an opto-electronic setup and that Laminar Chaos can be observed in such systems.
This demonstrates that Laminar Chaos is a robust phenomenon which can be observed in experiments.

\begin{figure}
	\includegraphics[width=0.45\textwidth]{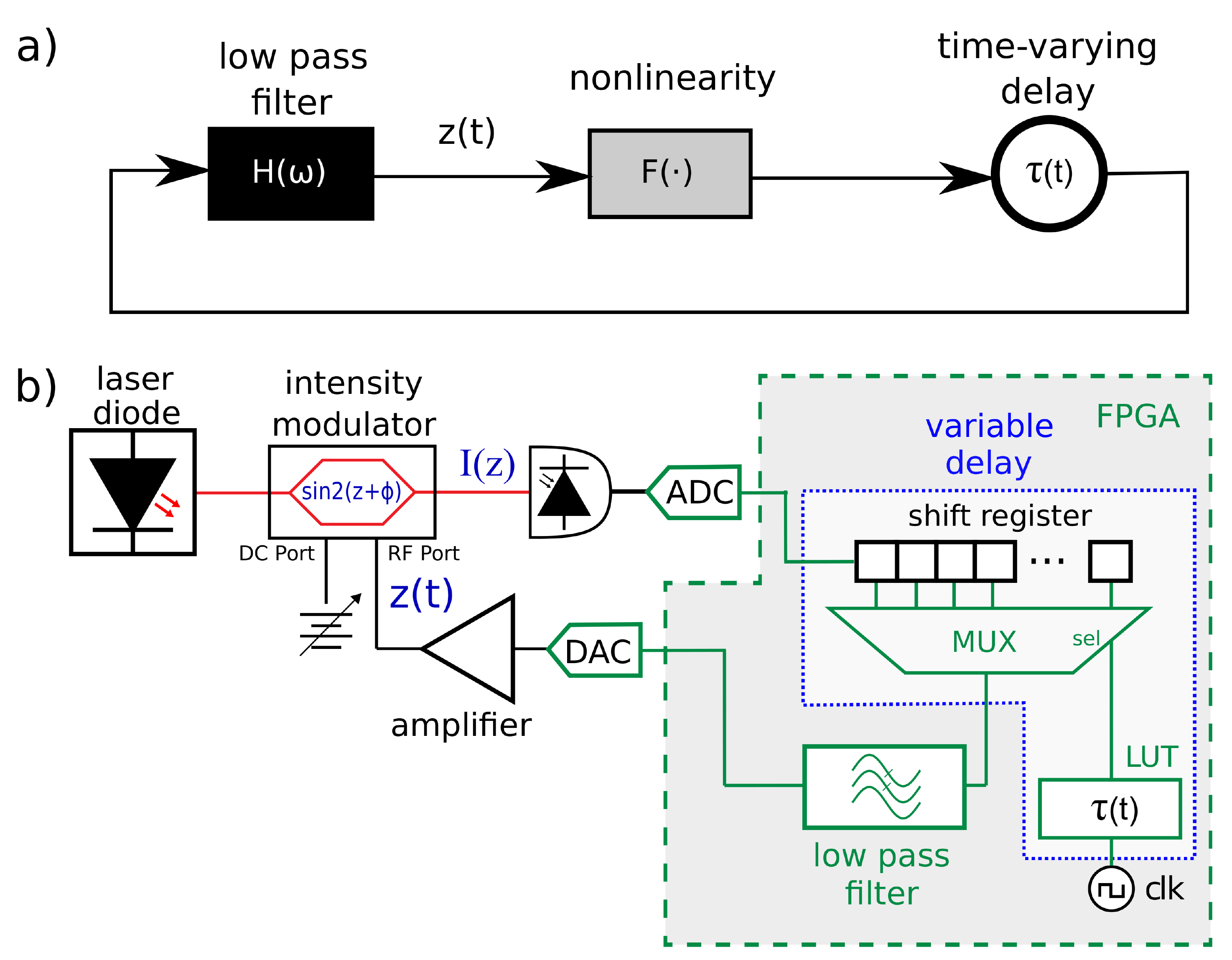}
	\caption{\label{fig:setup}
		(a) Block diagram of Eq. \ref{eq:sysdef}.
		This block diagram shows the essential ingredients for Laminar Chaos: a feedback loop with a band-limiting element (low pass filter), a nonlinearity, and a time-varying time delay.
		An additional condition is given by Eq. \ref{eq:crit}, which is visualized in Fig.~\ref{fig:crit}.
		(b) An illustration of the opto-electronic oscillator we used to observe Laminar Chaos.
		Red lines indicate the optical path, black lines indicate the electronic path, and green lines indicate signal-processing on the FPGA.
	}
\end{figure}

In Fig.~\ref{fig:setup}(a) the essential ingredients that need to be implemented to observe Laminar Chaos experimentally are visualized: a band-limited system with a nonlinear time-delayed feedback with time-varying delay.
This means that the output of the system at time $t$ depends on the past output of the system at time $t-\tau(t)$, which is fed back into the system after a nonlinear function $f$ is applied, and the system reacts to the feedback with a finite relaxation time.
Our opto-electronic setup is illustrated in Fig.~\ref{fig:setup}(b).
The nonlinearity $f(z)$ of the feedback is implemented by an integrated Mach-Zehnder intensity modulator and is given by
\begin{equation}
f(z)=\sin^2(z+\phi_0).
\end{equation}
In this case, the tunable parameter $\phi_0$ gives the bias point of the interferometer. The optical output of the modulator is converted to an electrical signal by a photoreceiver. This electrical signal is detected by an analog-to-digital converter (ADC), then low-pass filtered and delayed in a field-programmable gate array (FPGA). The FPGA has the capability of implementing a time-varying time delay via a tapped shift register and multiplexer (MUX). The delayed and filtered electrical signal is output by a digital-to-analog converter (DAC). The feedback loop is closed by applying the amplified DAC output to the RF port of the modulator. The parameter $\mu$ in Eq. \ref{eq:sysdef} is determined by the roundtrip gain of the feedback loop.

\begin{figure}[!ht]
	\includegraphics[width=0.39\textwidth]{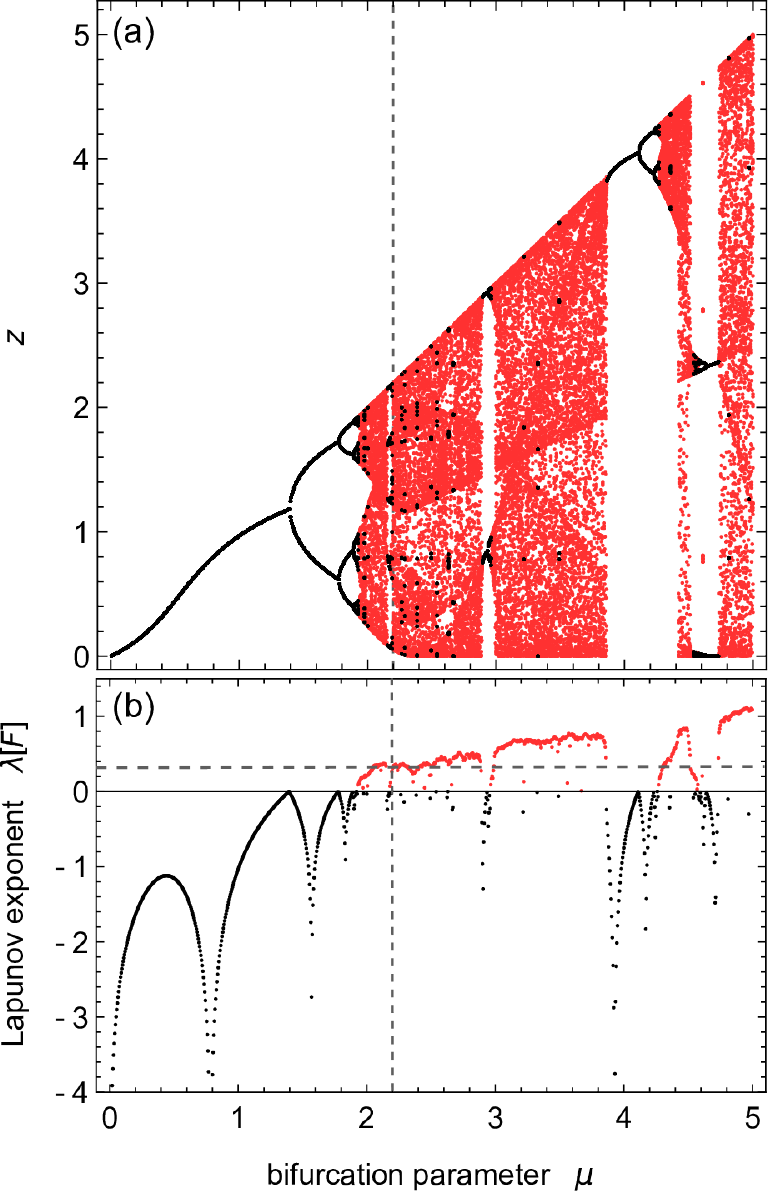}
	\caption{\label{fig:fmap}
		(a) Bifurcation diagram and (b) Lyapunov exponent $\lambda[F]$ of the map $z'=\mu\,f(z)=\mu \sin^2(z+\phi_0)$, where $\phi_0=\frac{\pi}{4}$, versus the bifurcation parameter $\mu$.
		A positive and negative Lyapunov exponent $\lambda[F]$ for a given $\mu$ is indicated by the usage of red and black color, respectively.
		The observations of Laminar Chaos presented in this paper were done with $\mu = 2.2$, which is indicated by the vertical dashed lines.
		In this case $\lambda[F] \approx 0.31$ (horizontal dashed line in (b)).
	}
\end{figure}

As shown in the previous section, the behavior of the map $z'=\mu f(z)$ must be chaotic for Laminar Chaos to be observed. In Fig~\ref{fig:fmap} a bifurcation diagram and the Lyapunov exponents of the corresponding map $z'=\mu\,f(z)$ under variation of $\mu$ are shown, where we chose $\phi_0=\frac{\pi}{4}$. For all work described here, we choose $\mu=2.2$, which corresponds to a chaotic region with $\lambda[F]\approx0.31$.

While the FPGA gives us great flexibility in the choice of the form of the time-varying delay $\tau(t)$, for all work presented here we consider a sinusoidal time-varying delay
\begin{equation}
\label{eq:vardelay}
\tau(t) = \tau_0 + \frac{A}{2\pi} \sin(2\pi\,t),
\end{equation}
where $\tau_0$ is the mean delay and $A$ is the delay amplitude, which are both measured in units of the delay period. The FPGA is clocked at a frequency $\nu_s=$100 kHz and the delay period $T_\tau=$ 10 ms, so that the period of the delay is divided into 1000 time steps. Additionally, the cutoff frequency of the low-pass filter $f_{\text{cutoff}}$=3183 Hz, so that $v_s\gg f_{\text{cutoff}}$. Therefore, the time discretization is sufficiently fine that this discrete time system is well modeled by Eqs. \eqref{eq:sysdef} and \eqref{eq:vardelay}. For a detailed presentation of a discrete time model of our opto-electronic oscillator and a discussion of the effects of the digitization of the delay, see Appendix A. A summary of the parameter values used in our experiment is given in Tab. I.

To analyze the robustness of Laminar Chaos to different amount of noise, we have implemented the possibility to add noise to the experiment in a controlled way. Specifically, at each time step we add numerically generated Gaussian white noise with zero
mean and standard deviation $\zeta$ to the normalized intensity
$I$ measured by the ADC. As we demonstrate in Appendix~\ref{sec:appnoise}, $\zeta=0.001$ is a good estimate for the inherent noise strength in our setup.  

\begin{figure}[t]
\includegraphics[width=0.4\textwidth]{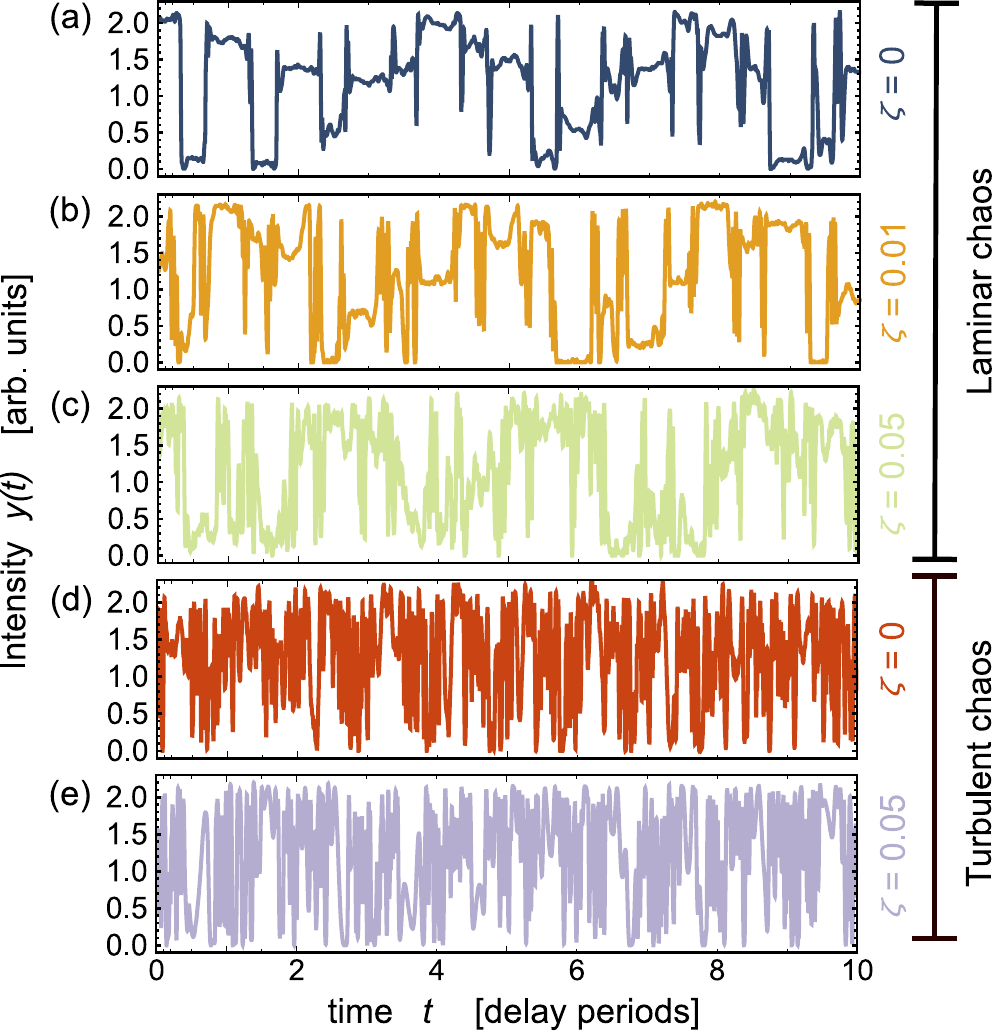}
\caption{\label{fig:trajexp}
Experimental time series, where $T=200$, $f(z)=\sin^2(t+\phi_0)$, $\mu=2.2$, $\phi_0=\frac{\pi}{4}$, and $\tau(t)=\tau_0 + 0.9\sin(2\pi\,t)/(2\pi)$ for different values of the mean delay $\tau_0$ and for different values of the strength $\zeta$ of the external noise.
The trajectories $(a)-(c)$ correspond to a dissipative delay ($\tau_0=1.5$) and show Laminar Chaos, whereas the trajectories (d) and (e) correspond to a conservative delay ($\tau_0=1.54$) and show Turbulent Chaos.
}
\end{figure}

Figure~\ref{fig:trajexp} shows exemplary trajectories, which were experimentally generated using the parameters in Tab.~\ref{tab:exppara} for different values of the strength $\zeta$ of the external noise.

\begin{table}[!ht]
\caption{Parameters of the experimental system and parameters that are normalized with respect to the delay period: delay period $T_\tau$, delay amplitude $A$, mean delay $\tau_0$, cutoff-frequency of the low pass filter $f_{\text{cutoff}}$, bifurcation parameter $\mu$, sampling frequency $\nu_s$}
\label{tab:exppara}
\begin{center}
\begin{tabular}{l|l|l}
parameter & normalized value & experimental value\\
\hline
\hline
$T_\tau$ & $1$ & $0.01\,\text{s}$\\
\hline
$A$ & $0.9$ & $0.009\,\text{s}$\\
\hline
$\tau_0$ & $1.5$, $1.54$ & $0.0150\,\text{s}$, $0.0154\,\text{s}$\\
\hline
$f_{\text{cutoff}}$ & $31.831$ & $3183\,\text{Hz}$\\
($T=200$) & & \\
\hline
$\mu$ & $2.2$ & $2.2$\\
\hline
$\nu_s$ & 1000 & $100\,\text{kHz}$
\end{tabular}
\end{center}

\end{table}

\section{A general model for Laminar Chaos in systems with noise}
\label{sec:model}

\begin{figure}[t]
\includegraphics[width=0.4\textwidth]{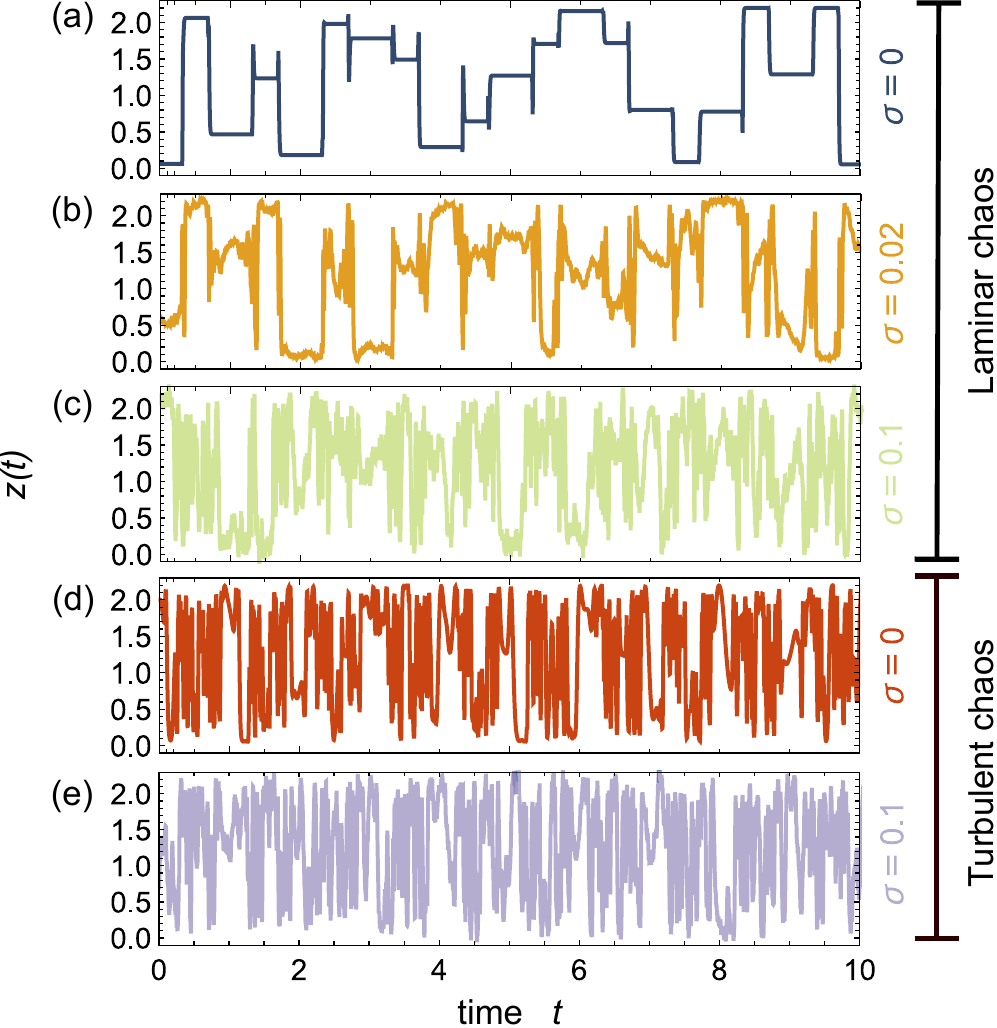}
\caption{\label{fig:traj}
Trajectories generated from Eq.~\eqref{eq:dimlessDDE}, where $T=200$, $f(z)=\sin^2(t+\phi_0)$, $\mu=2.2$, $\phi_0=\frac{\pi}{4}$, and $\tau(t)=\tau_0 + 0.9\sin(2\pi\,t)/(2\pi)$ for different values of the mean delay $\tau_0$ and the noise strength $\sigma$.
The trajectories $(a)-(c)$ correspond to a dissipative delay ($\tau_0=1.5$) and show Laminar Chaos, whereas the trajectories (d) and (e) correspond to a conservative delay ($\tau_0=1.54$) and show Turbulent Chaos.
}
\end{figure}

To understand the influence of noise on Laminar Chaos in our experiment and more general systems, and to benchmark the toolbox for the detection of Laminar Chaos that is provided in Section~\ref{sec:lamchaosnoise}, we derive and analyze in this section a nonlinear Langevin equation as a theoretical model.
Laminar Chaos was originally found in the first order DDE~\eqref{eq:sysdef}, but our experimental system is a second order system, due to the second order Butterworth filter that is used in our setup (cf. Appendix~\ref{sec:appnoise}).
To obtain a general model, we consider systems of arbitrary order $D$.
Taking into account the inherent noise of experiments by an additive noise term, we consider the general $D$th-order stochastic DDE
\begin{equation}
\label{eq:ndimsysdef1}
\mathcal{D}[x](\tilde{t})=\sum_{d=0}^D a_d \, x^{(d)}(\tilde{t}) = \tilde{\mu}\,f(x(\tilde{R}(\tilde{t}))) + \tilde{\sigma}\,\tilde{\xi}(\tilde{t}) = I(\tilde{t}),
\end{equation}
where $\tilde{R}(\tilde{t})=\tilde{t}-\tilde{\tau}(\tilde{t})$, $\tilde{\tau}(\tilde{t}+T_\tau)=\tilde{\tau}(\tilde{t})$, $x^{(d)}(\tilde{t})$ denotes the $d$th derivative of $x(\tilde{t})$, and $\tilde{\xi}(\tilde{t})$ is Gaussian white noise with $\langle \tilde{\xi}(\tilde{t}) \rangle = 0$ and $\langle \tilde{\xi}(\tilde{t})\tilde{\xi}(\tilde{t}') \rangle=\delta(\tilde{t}-\tilde{t}')$.
The essential ingredients for Laminar Chaos visualized in Fig.~\ref{fig:setup}(a), namely, the band limiting element and the nonlinear feedback with variable delay are represented by the terms $\mathcal{D}[x](\tilde{t})$ and $\tilde{\mu}\,f(x(\tilde{R}(\tilde{t})))$, respectively.
In detail, a solution $x(\tilde{t})$ of Eq.~\eqref{eq:ndimsysdef1} can be interpreted as a filtered version of the input signal $I(\tilde{t})$, i.e., $x(\tilde{t})$ is generated by applying the inverse $\mathcal{D}^{-1}$ of the differential operator to $I(\tilde{t})$.
For suitable parameters (for example, $D=1$, $a_0=a_1=1$ as in Eq.\eqref{eq:sysdef}), the operator $\mathcal{D}^{-1}$ is a smoothing operator and acts as a low pass filter.
This means that $\mathcal{D}^{-1}[I](\tilde{t})=I(\tilde{t})$ if $I(\tilde{t})$ is constant, allowing laminar phases to develop if certain conditions are fulfilled.
The general analysis of the question which parameters lead to smoothing operators that allow the development of Laminar Chaos goes beyond the scope of this paper and will be discussed elsewhere.
However, it is clear that a bounded input $I(\tilde{t})$ leads to a bounded output $x(\tilde{t})$ only if the characteristic roots of $\mathcal{D}[x](\tilde{t})=0$, which are the poles of the transfer function of the filter, have a negative or vanishing real part.

By introducing a dimensionless time $t$ via $\tilde{t}=T_\tau\,t$ and the new variable $z(\tilde{t}/T_\tau)=x(\tilde{t})$, and with $\tilde{\xi}(t\,T_\tau)=\xi(t)/\sqrt{T_\tau}$ we obtain
\begin{equation}
\label{eq:ndimsysdef2}
\sum_{d=0}^D \frac{a_d}{T_\tau^d} \, z^{(d)}(t) = \tilde{\mu}\,f(z(R(t))) + \frac{\tilde{\sigma}}{\sqrt{T_\tau}}\,\xi(t),
\end{equation}
where $R(t)=\tilde{R}(t\,T_\tau)/T_\tau=R(t+1)-1$ such that $\tau(t+1)=\tau(t)$.
For $T_\tau \gg \tau(t)$, which corresponds to a large delay compared to the internal time scale of the system, we neglect terms with higher order in $T_\tau^{-1}$ and obtain
\begin{equation}
\frac{1}{T_\tau}\,a_1 \dot{z}(t) + a_0 z(t) = \tilde{\mu}\,f(z(R(t))) + \frac{\tilde{\sigma}}{\sqrt{T_\tau}}\,\xi(t).
\end{equation}
Dividing by $a_0$ and substituting $T=(a_0/a_1)T_\tau$, $\mu=\tilde{\mu}/a_0$, $\sigma=\tilde{\sigma}/\sqrt{a_0 a_1}$ leads to the dimensionless DDE
\begin{equation}
\label{eq:dimlessDDE}
\frac{1}{T} \dot{z}(t) + z(t) =  \mu \, f(z(R(t))) + \frac{\sigma}{\sqrt{T}}\,\xi(t).
\end{equation}
By the assumption that $a_0$ and $a_1$ have the same sign, we ensure $T>0$.
This means that the ODE part of Eq.~\eqref{eq:dimlessDDE}, which is given by its left hand side, is stable.

We see that in the large delay limit the $D$th order DDE~\eqref{eq:ndimsysdef1} can be suitably approximated by the first order DDE~\eqref{eq:dimlessDDE}, which is used for the further theoretical analysis.
Exemplary trajectories of Eq.~\eqref{eq:dimlessDDE} are shown in Fig.~\ref{fig:traj}.

\section{How to detect Laminar Chaos}
\label{sec:lamchaosnoise}

In this section we describe distinctive features of Laminar Chaos that are robust against experimental noise and measurement noise.
We provide a toolbox for the detection of these features, which allows the identification of Laminar Chaos in experimental data.
If a trajectory shows Laminar Chaos, the toolbox provides valuable information about the considered system, such as the nonlinearity $\mu\,f(z)$ of the delayed feedback, the delay period $T_\tau$, and the rotation number $\rho$ of the access map that is associated with the time-varying delay.
The following distinctive features are considered: the roughly periodic structure of the derivative of the trajectories due to the periodic sequence of durations of the laminar phases and the one-dimensional map that connects the intensity levels of the laminar phases.
The characterization of a signal showing Laminar Chaos can be divided into two steps.
Firstly, one has to verify the existence of laminar phases, where the duration of the phases follows a periodic sequence with the same period $T_\tau$ as the delay.
A method for doing this, which also enables the detection of the position of the laminar phases, is introduced in Sec.~\ref{sec:detect_lamphases}.
The second step of the verification is performed in Sec.~\ref{sec:nlinrec}, where it is verified that the intensity levels are connected by a one-dimensional map.

In Sec.~\ref{sec:scanlamchaos} we consider the autocorrelation function and a related correlation length, which can be useful to scan for parameters that correspond to Laminar Chaos, if it is known that the considered system can show Laminar Chaos for certain parameters.

\begin{figure}[!t]
\includegraphics[width=0.4\textwidth]{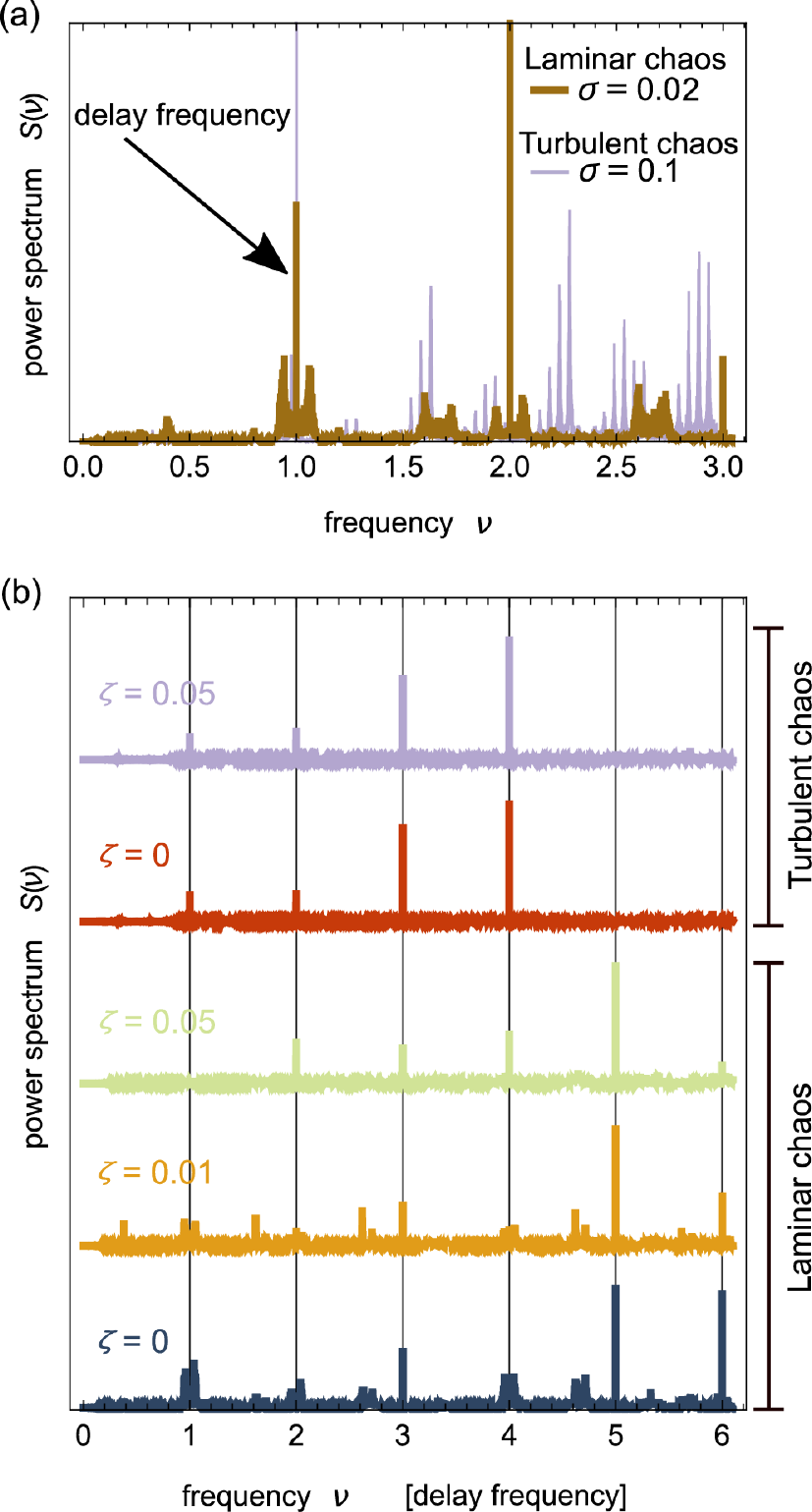}
\caption{\label{fig:pws}
Estimation of the delay period:
(a) Low frequency region of the power spectrum of the approximated derivatives ($h \approx 0.033$) of trajectories of Eq.~\eqref{eq:dimlessDDE}, where $T=200$, $f(z)=\sin^2(t+\phi_0)$, $\mu=2.2$, $\phi_0=\frac{\pi}{4}$, and $\tau(t)=\tau_0 + 0.9\sin(2\pi\,t)/(2\pi)$, which show Laminar Chaos ($\tau_0=1.5$, $\sigma=0.02$) and Turbulent Chaos ($\tau_0=1.54$, $\sigma=0.1$).
The arrow points to the peak, which corresponds to the delay period and can be easily identified.
The spectra are computed by averaging 10 spectra, which are in turn computed from distinct windows of the length of $10^5$ delay periods.
(b) Power spectra of the approximated derivatives ($h = 0.001\,T_\tau$) of the experimental trajectories that are partially shown in Fig.~\ref{fig:trajexp}, where the frequency is given in multiples of the delay frequency $\nu_\tau=T_\tau^{-1}$.
Some of the peaks at the multiples of the delay frequency $\nu_\tau$ (vertical lines) are clearly visible. However, if the delay frequency $\nu_\tau$ is not known, longer time series are needed to distinguish the peaks at the multiples of $\nu_\tau$ from the background as it is possible for the spectra in (a).
}
\end{figure}

\subsection{Detection of the laminar phases}
\label{sec:detect_lamphases}

\begin{figure}[!t]
\includegraphics[width=0.4\textwidth]{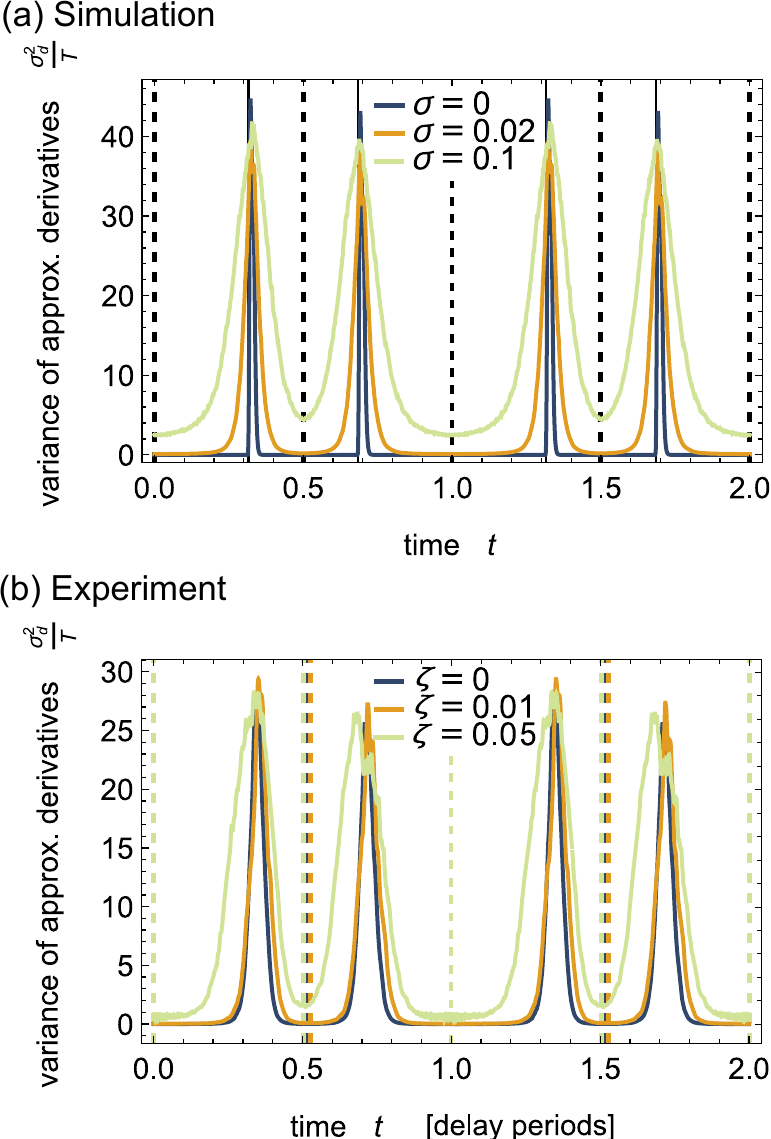}
\caption{\label{fig:tempvard}
Detection of the laminar phases:
Temporal distribution of the variance $\sigma_d^2$ (in units of $T$) of the approximated derivatives ($h \approx 0.0033$) of
(a)
laminar chaotic trajectories of Eq.~\eqref{eq:dimlessDDE} with the same parameters as in Fig.~\ref{fig:traj} and
(b)
experimental trajectories for different external noise strengths $\zeta$.
$\sigma_d^2$ is defined by Eq.~\eqref{eq:tdistvar} and is a measure for the fluctuation strength of the trajectory at a specific time relative to the internal clock induced by the time-varying delay.
The laminar phases are located in the neighborhood of the attractive fixed points of the reduced access map (dashed lines in (a)) and the burst-like transitions between the laminar phases are located at the repulsive fixed points of the access map (solid lines).
In this case the denominator of the rotation number of the access map is given by $q=2$, which leads to two laminar phases per period.
The temporal distribution of the variance $\sigma_d^2$ in (b) for the experimental time series was shifted, such that the numerically computed local minimum that corresponds to the longest laminar phase is located at $t=0$.
The dashed lines indicate the local minima, which where determined numerically.
}
\end{figure}

\begin{figure}[!t]
\includegraphics[width=0.4\textwidth]{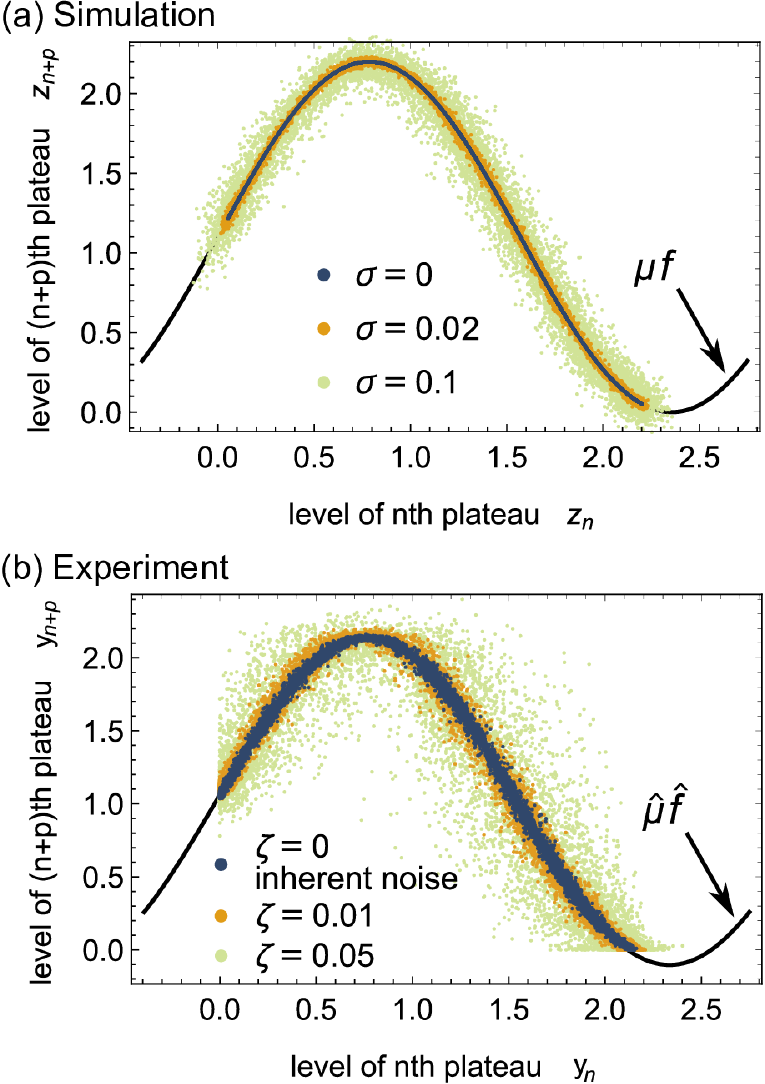}
\caption{\label{fig:nlinrec}
(a) Reconstruction of the nonlinearity (solid black line) of Eq.~\eqref{eq:dimlessDDE} from a trajectory that shows Laminar Chaos by plotting the intensity level of the $(n+p')$th laminar phase $z_{n+p'}$ against the intensity level of the $n$th laminar phase $z_n$.
In the case of Laminar Chaos, there are infinitely many positive $p'\in\mathbb{N}$ such that the points $(z_{n}, z_{n+p'})$ resemble a line.
For the smallest of these numbers $p'=p$ this line is the nonlinearity $\mu\,f(z)$ (plotted here).
We used the same parameters as in Fig.~\ref{fig:traj}.
(b) Reconstruction of the nonlinearity for experimental trajectories, which where realized with external noise of different strength $\zeta$.
The black line represents the fit $\hat{\mu}\hat{f}(y)$ of the reconstructed nonlinearity for $\zeta=0$, where $\hat{\mu} \approx 2.229 \pm 0.002$ and $\hat{f}(y)=\sin^2(y+\hat{\phi}_0)+\hat{c}_0$, where $\hat{\phi}_0 \approx 0.8074 \pm 0.0004$ and $\hat{c}_0 \approx -0.103 \pm 0.002$.
The error estimates represent the confidence intervals returned by \texttt{NonlinearModelFit} in \emph{Mathematica} 11.3 (twice the standard error).
The clipping below $y_n=0$ and $y_{n+p}=0$ is due to the fact that the system is an optical system, where the amplitude of the trajectories corresponds to the light intensity, which can not be lower than zero.
}
\end{figure}

If the trajectory shows Laminar Chaos, its derivative shows the following behavior.
Without noise, which means $\sigma=0$, the derivative is roughly zero between the bursts, i.e., it is characterized by phases with approximately zero amplitude, which are periodically interrupted by short large amplitude bursts.
In the presence of noise we consider the approximated derivative
\begin{equation}
\label{eq:approxderiv}
\Delta_h[z](t) = \frac{z(t+h)-z(t)}{h}
\end{equation}
instead of the derivative, since the latter is not well defined.
In this case the approximated derivative is characterized by phases of small amplitude which are periodically interrupted by short large amplitude bursts.
The periodic part of the derivative of the trajectory leads to a large peak in the power spectrum at the frequency of the delay $\nu_\tau$ as highlighted by the black arrow in Fig.~\ref{fig:pws}(a), where the power spectrum of the approximated derivative $\Delta_h[z](t)$ is plotted.
For Turbulent Chaos, the power spectrum also shows a large peak at the delay frequency $\nu_\tau$, since, in this case, the time-varying delay leads to a quasiperiodic modulation of the signal with two frequencies, where the  delay frequency is one of them (cf. non-resonant doppler effect in Ref.~\cite{muller_resonant_2019}).

To determine the position of the laminar phases of a laminar chaotic trajectory $z(t)$, we consider the temporal distribution of the variance $\sigma_d^2[z](t)$ of $\Delta_h[z](t)$, which is defined by
\begin{equation}
\label{eq:tdistvar}
\sigma_d^2[z](t) =  \lim_{N\to\infty} \frac{1}{N} \sum_{n=0}^{N-1} \left( \Delta_h[z](t+n) \right)^2 - (\mu_d[z](t))^2,
\end{equation}
where
\begin{equation}
\mu_d[z](t) = \lim_{N\to\infty} \frac{1}{N} \sum_{n=0}^{N-1} \Delta_h[z](t+n),
\end{equation}
and $\Delta_h[z](t)$ is defined by Eq.~\eqref{eq:approxderiv}.
So the average is taken over equidistant times $t'$ starting with $t'=t$, where the distance between them is the period of the time-varying delay, which equals one, since the time $t$ is measured in delay periods.
If the delay period $T_\tau$ is unknown for an experimental time series, it can be determined by analyzing the power spectrum as shown in Fig.~\ref{fig:pws}.
The quantity $\sigma_d^2[z](t)$ is very sensitive to errors of the delay period, since large multiples of the delay period occur due to the sampling of the trajectory at multiples of the delay period.
So the delay period must be determined very accurately to compute a reasonable approximation of $\sigma_d^2[z](t)$.
However, this sensitivity can be exploited to determine the delay period very accurately as shown in Appendix~\ref{sec:apx_delayperiod}.
In Fig.~\ref{fig:tempvard} $\sigma_d^2$ is shown for exemplary trajectories, which show Laminar Chaos and are generated for different noise strengths.
If a trajectory is characterized by periodically alternating high and low frequency phases and the period equals the delay period as in the case of Laminar Chaos, $\sigma_d^2$ alternates periodically between high and low values, which correspond to the high and low frequency phases of the trajectory.
For increasing noise strength $\sigma$, the fluctuation strength in the low amplitude phases increases, such that the periodic structure is still present but gets blurred as visualized in Fig.~\ref{fig:tempvard}.
The local minima of $\sigma_d^2$ can be used to determine the position of the laminar phases of laminar chaotic trajectories.
The denominator $q$ of the rotation number of the access map equals the number of laminar phases per period and, thus, it can be also determined by this method.

\subsection{Reconstruction of the nonlinearity}
\label{sec:nlinrec}

To check whether the intensity levels of the detected laminar phases are connected by a one-dimensional map, we plot the intensity level of the $(n+p')$th laminar phase $z_{n+p'}$ against the intensity level of $n$th laminar phase $z_n$, where $p'\in\mathbb{N}$ and $p'>0$.
For our analysis, the intensity levels of the laminar phases are the values of the trajectory at the local minima of the temporal distribution of the variance $\sigma_d^2$, which is visualized in Fig.~\ref{fig:tempvard}.
In the case of Laminar Chaos, for the correct $p'=p$ the points $(z_{n}, z_{n+k\,p})$ resemble the graph $(z,(\mu\,f)^k(z))$.
So $p'$ was chosen correctly, if $p'$ is the smallest number such that the points $(z_{n}, z_{n+p'})$ resemble a line, as shown in Fig.~\ref{fig:nlinrec}.
If $p'$ is wrongly chosen, then the points $(z_{n}, z_{n+p'})$ fill a set $\mathcal{S}$, which is the union of Cartesian products of pairs of the chaotic bands of the map, since in this case $z_{n+p'}$ and $z_{n}$ are independent.
The set $\mathcal{S}$ is two-dimensional except for cases, where the dimension of the attractor is smaller than one, which requires $\lambda[F] \leq 0$.
The number $p$ denotes the numerator of the rotation number $\rho=-\frac{p}{q}$ of the access map.
Thus, the latter can also be reconstructed from laminar chaotic trajectories.
If no such $p$ can be found, the reconstruction of the nonlinearity fails as demonstrated in Appendix~\ref{sec:apprecturb} and the trajectory can not be characterized as Laminar Chaos.

\subsection{How to scan parameters for Laminar Chaos}
\label{sec:scanlamchaos}

\begin{figure}[t]
\includegraphics[width=0.4\textwidth]{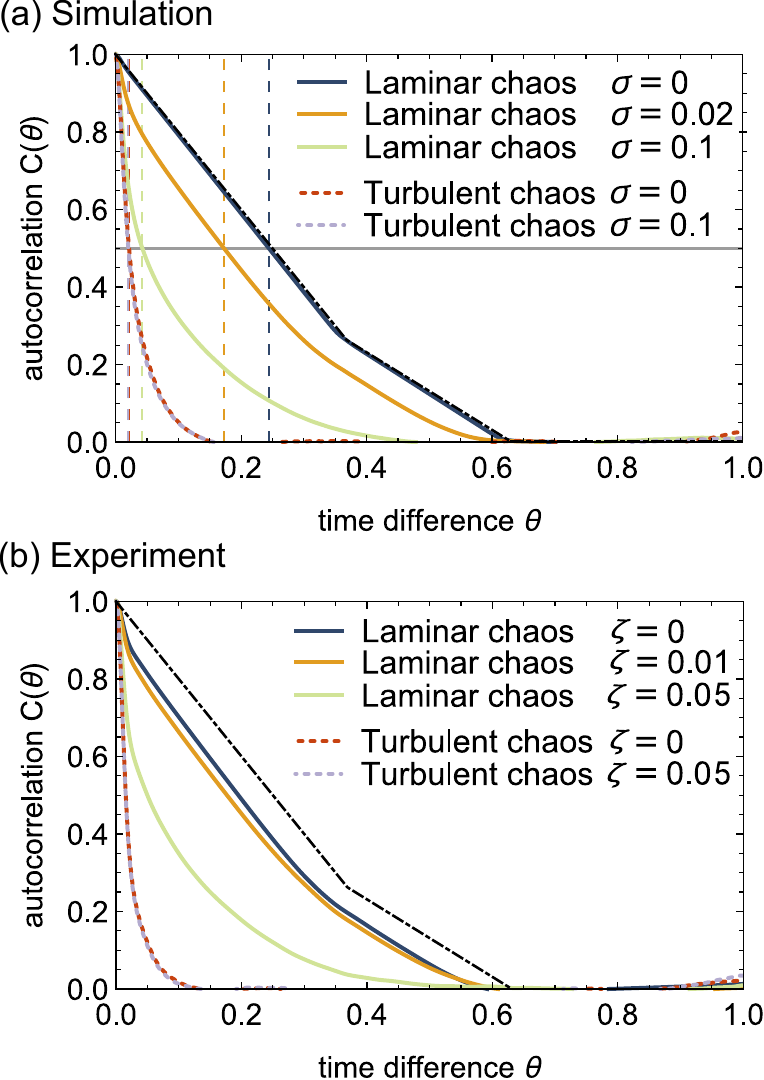}
\caption{\label{fig:corr}
(a) Autocorrelation $C(\theta)$ of trajectories of Eq.~\eqref{eq:dimlessDDE} with the same parameters as in Fig.~\ref{fig:traj}.
The time-average in Eq.~\eqref{eq:autocorr} was taken over $10000$ delay periods.
As a possible quantity for the distinction between laminar and Turbulent Chaos, the half of the full width at half maximum is indicated by the vertical dashed lines.
For $\sigma=0$ the autocorrelation of the laminar chaotic trajectory is approximately piecewise linear as indicated by Eq.~\eqref{eq:autocorr_lamchaos} (unevenly dashed black line).
Here we have $q=2$, leading to an autocorrelation function with two linear segments with non-zero slope.
The values of $\theta$ where the slope changes correspond to the widths of the two laminar phases per delay period, which are obtained for the chosen set of parameters.
(b) Autocorrelation $C(\theta)$ of experimental trajectories, which where realized with external noise of different strengths $\zeta$.
The time-average in Eq.~\eqref{eq:autocorr} was taken over at least $5000$ delay periods.
Since noise is always inherent to experimental systems in the absence of external noise ($\zeta=0$), the autocorrelation $C(\theta)$ in this case also slightly deviates from the theoretical behavior given by Eq.~\eqref{eq:autocorr_lamchaos} (unevenly dashed black line).
}
\end{figure}

\begin{figure}[t]
\includegraphics[width=0.4\textwidth]{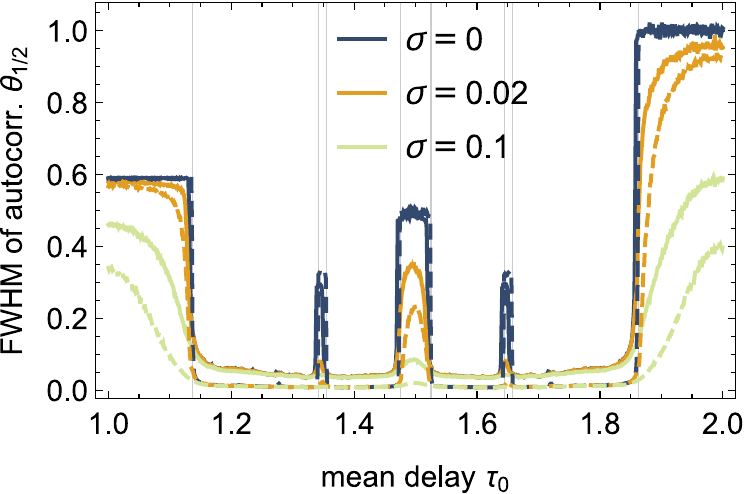}
\caption{\label{fig:corrfwhm}
How to find delay parameters that correspond to Laminar Chaos: Full width at half maximum (FWHM) $\theta_{1/2}$ of the autocorrelation $C(\theta)$ of trajectories of Eq.~\eqref{eq:dimlessDDE}, $f(z)=\sin^2(t+\phi_0)$, $\mu=2.2$, $\phi_0=\frac{\pi}{4}$, and $\tau(t)=\tau_0 + 0.9\sin(2\pi\,t)/(2\pi)$ under variation of $\tau_0$ for different noise strengths $\sigma$, with $T=200$ (solid) and $T=1000$ (dashed).
For Laminar Chaos the FWHM is large, whereas for Turbulent Chaos the FWHM is small.
The boundaries of the regions of $\tau_0$ where the criterion~\eqref{eq:crit} for Laminar Chaos is fulfilled are represented by the vertical lines.
The slight deviation between the latter and the boundaries represented by the jumps of the FWHM for $\sigma = 0$ is caused by the finite $T<\infty$ and vanishes for $T\to\infty$.
}
\end{figure}

Laminar Chaos is characterized by two time-scales, one of the order of $\frac{1}{T}$ (small for large $T$) and one of the order of the duration of the laminar phases ($\frac{1}{q}$, independent of $T$).
This means that points that are close in time are strongly correlated even if the system is able to relax very fast with the relaxation rate $T$, i.e., it is able to admit high frequencies.
If we analyze the output of a system that can show only Laminar Chaos or Turbulent Chaos, this fact can be used to distinguish between the two by the autocorrelation of the time series.
The autocorrelation function is defined by
\begin{equation}
\label{eq:autocorr}
C(\theta) = \lim_{t\to\infty} \frac{\langle z(t') z(t'+\theta) \rangle_t-\langle (z(t')) \rangle_t^2}{\langle (z(t'))^2 \rangle_t - \langle (z(t')) \rangle_t^2},
\end{equation}
where the time-average is given by
\begin{equation}
\langle \;\cdot\; \rangle_t = \frac{1}{t} \int_{0}^{t} \;\cdot\; dt' .
\end{equation}

The autocorrelation functions of trajectories of Eq.~\eqref{eq:dimlessDDE} with the same parameters as in Fig.~\ref{fig:traj} are shown in Fig.~\ref{fig:corr}(a).
Increasing the noise strength $\sigma$ leads to a faster decay of correlations, since the laminar phases are more and more distorted as it is visible in Fig.~\ref{fig:traj}.
However, even for comparably large noise strengths the decay of $C(\theta)$ is slower than for Turbulent Chaos.
For the latter in Fig.~\ref{fig:corr}(a) the influence of the noise is not visible, since correlations decay already very fast in the absence of noise, indicating the high-dimensional nature of the related chaotic dynamics.
These results can be experimentally reproduced as shown in Fig.~\ref{fig:corr}(b), where the autocorrelation function $C(\theta)$ was computed from experimental time series for the same parameters as in Fig.~\ref{fig:trajexp}.

As a numerical indicator for Laminar or Turbulent Chaos we consider the full width at half maximum $\theta_{1/2}$ of $C(\theta)$.
For Turbulent Chaos with only one chaotic band, $\theta_{1/2}$ is of the order of $\frac{1}{T}$ and, contrarily, for Laminar Chaos $\theta_{1/2}$ is at least of the order of $1/q$, which is of the order of the duration of the laminar phases and, thus, it is much larger than $1/T$.
Consequently, a good threshold for $\theta_{1/2}$ above which the trajectory can be classified as Laminar Chaos should be of the order of $\frac{1}{T}$ or larger.
In Fig.~\ref{fig:corrfwhm} $\theta_{1/2}$ is plotted as a function of the mean delay $\tau_0$, where all other parameters are kept constant.
Particularly, we choose a delay amplitude $A=0.9$, which corresponds to the dashed vertical line in Fig.~\ref{fig:crit}.
If the parameter $T$ is unknown but considered to be large, it can be measured.
For example, one can prepare the system with a constant initial function, which leads (for large $T$) to the relaxation to an almost constant state in the following time-interval.
The related relaxation time in the corresponding dimensionless system is $\frac{1}{T}$.

So we have seen, that the FWHM of the autocorrelation function is a simple criterion to distinguish Laminar and Turbulent Chaos.
In general and especially when the generating system is completely unknown, the FWHM of the autocorrelation function alone is not sufficient to detect Laminar Chaos but it can serve as a first indicator.
If the considered system shows chaos that is characterized by multiple chaotic bands, Laminar and Turbulent chaos are much harder to distinguish by using the autocorrelation only.
In this case, the trajectory alternates periodically between the chaotic bands, which leads to a large contribution to the autocorrelation for both types of chaos.
To use the autocorrelation as an indicator anyway, this contribution must be removed, which can be achieved, for example, by preprocessing the trajectories such that the periodic alternation between the chaotic bands is subtracted.

As illustrated in Fig.~\ref{fig:corr}, the autocorrelation function $C(\theta)$ is a nearly piecewise linear function for vanishing noise strengths $\sigma$ and $\zeta$.
To give an explanation for this, we derive an expression for $C(\theta)$ in the absence of noise, which is exact in the limit $T\to\infty$.
To obtain an expression for $C(\theta)$ that is independent of the nonlinearity of the feedback $\mu f(z)$, we bound the modulus of the time variable $\theta$ by the delay period, i.e., $|\theta| \in [0,1]$, and assume that $p>q$.
The integers $p$ and $q$ are the numerator and the denominator of the rotation number $\rho=-\frac{p}{q}$ of the access map given by Eq.~\eqref{eq:accessmap}.
The integers $p$ and $q$ are also the numbers of laminar phases per state interval $\mathcal{I}_n$ and per delay period, respectively.
We denote the encountered durations of the laminar phases by $d_j$ and put them in ascending order, i.e. $d_1 < d_2 < \dots < d_q$, to get a simple expression for $C(\theta)$.
The $d_j$ are measured in delay periods, with the result that $\sum_{j=1}^q d_j = 1$, since we have $q$ durations $d_j$ per delay period.
If the dynamics of the DDE system is characterized by only one chaotic band, we obtain for the $k$th linear segment of the autocorrelation function
\begin{equation}
\label{eq:autocorr_lamchaos}
C(\theta) = \sum_{j=k}^q (d_j - |\theta|), \quad \text{with } |\theta| \in [d_{k-1},d_k],
\end{equation}
where $k\in\{1,2,\dots q\}$, $d_0=0$, and $C(\theta)=0$, if $|\theta| \in [d_q,1]$.
Here we use the fact that, given a fixed $\theta$, the time-average for the covariance of $z(t')$ and $z(t'+\theta)$ consists of two parts.
The first part is obtained by averaging only over values of $t'$ for which $t'$ and $t'+\theta$ do not belong to the same laminar phase.
The laminar phases inside a time interval of length of one delay period are pairwise independent for $p>q$.
This is because the laminar phases inside the state interval are pairwise independent for $p>1$ as pointed out in Ref.~\cite{muller_resonant_2019}, and for $p>q$ the state interval is longer than the delay period.
As a result, the first part vanishes.
The remaining part equals the variance of $z(t)$, since $t'$ and $t'+\theta$ belong to the same laminar phase, $z(t)$ consists of exactly constant laminar phases in the limit $T\to\infty$, and thus $z(t')=z(t'+\theta)$.
Consequently, the laminar phase of length $d_j$ contributes to $C(\theta)$ only if $|\theta| \leq d_j$ and the contribution is given by $d_j-|\theta|$.
For $\theta \in [d_{k-1},d_k]$ the laminar phases of length $d_k,d_{k+1},\dots,d_q$ contribute, which results in Eq.~\eqref{eq:autocorr_lamchaos}.
For $p \leq q$ the autocorrelation function is also piecewise linear but the slope of the linear segments differ from Eq.~\eqref{eq:autocorr_lamchaos}, since certain laminar phases inside a time interval of length of one delay period are connected by the map $z'=\mu\,f(z)$.
This leads to an additional contribution to the part of the covariance of $z(t')$ and $z(t'+\theta)$, where $t'$ and $t'+\theta$ do not belong to the same laminar phase.

For the examples in Fig.~\ref{fig:corr} that show Laminar Chaos, we have chosen parameters such that we have $q=2$.
In this case Eq.~\eqref{eq:autocorr_lamchaos} results in
\begin{equation}
C(\theta) =
\begin{cases}
\sum_{j=1}^2 (d_j - |\theta|) = 1-2|\theta|, & \text{if } |\theta| \in [0,d_1]\\
\sum_{j=2}^2 (d_j - |\theta|) = d_2-|\theta|, & \text{if } |\theta| \in [d_1,d_2]\\
0, & \text{if } |\theta| \in [d_2,1].
\end{cases}
\end{equation}
This means that in the ideal case without noise and in the limit $T\to\infty$ the autocorrelation consists of two linear segments with non-zero slope for $\theta\in[0,1]$.
In the presence of noise the shape of $C(\theta)$ is smoothed with the results that the piecewise linear character gets lost.
Due to the inherent noise of the experiment, even in the absence of external noise ($\zeta=0$) the autocorrelation deviates from Eq.~\eqref{eq:autocorr_lamchaos}.

\section{Conclusion}

We have demonstrated that Laminar Chaos is a robust phenomenon, which can be observed in experimental systems.
Based on the analysis of the dynamics of an opto-electronic setup and the corresponding model given by a $D$th order nonlinear delayed Langevin equation, we have demonstrated that Laminar Chaos is characterized by robust features, which survive the presence of noise.
This means that these features can be detected in noisy experimental time series without any knowledge about the system that has generated these time series.
In fact, during the detection procedure several details of the generating system, such as the nonlinearity of the feedback and certain dynamical quantities of the access map, can be determined.
We have shown that a time series showing Laminar Chaos is characterized by a periodic sequence of regions that are characterized by low and high amplitude fluctuations.
By the temporal distribution of the variance of the approximated derivatives, which is given by Eq.~\eqref{eq:tdistvar}, we provide a tool to detect the low and high frequency regions, which can also be exploited for an accurate determination of the period of the delay, which is in general unknown.
For Laminar Chaos, in the absence of noise, the intensity levels of the laminar phases are connected by a one-dimensional map, which is given by the nonlinearity of the feedback.
This feature survives the presence of noise in the sense that the intensity levels of the regions with low fluctuation amplitude can be used to reconstruct the nonlinearity of the feedback from time series that show Laminar Chaos.
In contrast, if the reconstruction fails, Laminar Chaos can be excluded and the trajectory shows, for example, Turbulent Chaos.
This classification is possible even in the case of comparably large noise strengths, where Laminar Chaos is visually indistinguishable from Turbulent Chaos.

Our experimental setup is a band limited system with nonlinear time-delayed feedback, where the delay is time-varying.
The delay and the low pass filter are implemented digitally by an FPGA, whereas the nonlinear feedback generated by the optical part of the system evolves in continuous time.
Thus, we have demonstrated that hybrid systems also can show Laminar Chaos.
Moreover, due to the fact that our experimental system is band limited by a second order Butterworth filter and with the theoretical analysis in Sec.~\ref{sec:model} it is now clear that Laminar Chaos is not restricted to first order systems.

Our results support the statements in Ref.~\cite{muller_laminar_2018,muller_resonant_2019}, where we have conjectured that the intensity levels of the laminar phases can be used to encode information for computational or cryptographic purposes.
For example, due to the robustness of the mapping between the intensity levels of the laminar phases, a chaos based logic based on the ideas behind Ref.~\cite{murali_logic_2009,ditto_chaogates:_2010,ditto_exploiting_2015} may be implemented easily by a system that exhibits Laminar Chaos.

During the review process Ref.~\cite{jungling_laminar_2020} was published, where Laminar Chaos was found in a nonlinear electronic circuit with delay clock modulation. This further verifies our result that Laminar Chaos is a robust phenomenon and it demonstrates that Laminar Chaos can be observed in a variety of systems.

\begin{acknowledgments}
The authors thank Don Schmadel and Thomas E. Murphy for helpful discussions.
D.MB., A.O., and G.R. acknowledge partial support from the German Research Foundation (DFG) under the grant no. 321138034.
This work was supported by ONR Grant No. N000141612481 (J.D.H. and R.R.).
\end{acknowledgments}

\appendix

\section{Experimental system description and noise strength estimation}
\label{sec:appnoise}

As stated in Sec.~\ref{sec:exprealize}, the FPGA is clocked and operates in discrete time. When the sampling rate of the FPGA is sufficiently faster than the cut-off frequency of the low-pass filter and the frequency of the periodically varying time delay, the system can be accurately approximated by Eq. (1). In some cases, such as when studying the impact of the discrete time on Laminar Chaos, it may be useful to have a discrete-time model of the system. We provide such a model in this section.
The equation for the digital low-pass filter, which generates a smoothed version of the normalized intensity $I$, is given by
\begin{eqnarray}
\label{eq:digfilter}
y[m] = &-& a_1 y[m-1] -a_2 y[m-2] \nonumber\\
&+& \mu \{b_0 I[m] +b_1 I[m-1] + b_2 I[m-2]\},
\end{eqnarray}
where $a_1=-1.71869074$, $a_2=0.75364692$, $b_0=0.00873905$, $b_1=0.01747809$, and $b_2=0.00873905$.
It represents a discrete time approximation of a second order Butterworth filter with gain $\mu$ and cutoff-frequency $f_{\text{cutoff}} = T/(2\pi) \approx 31.831$, which can be derived from the transfer function via the bilinear transform \cite{smith_scientist_1997}.
The signal $I$ that is fed into the low pass filter is the normalized output of the Mach-Zehnder intensity modulator, that is
\begin{equation}
\label{eq:mzmod}
I[m]=f( y[m-\tau_s[m]] ).
\end{equation}
The input of the Mach-Zehnder intensity modulator at time $t$ is given by the output $y$ of the filter at time $t-\tau(t)$, where the FPGA realizes a discrete time version of the delay, which is given by
\begin{equation}
\label{eq:digdelay}
\tau_s[m] = \left\lfloor m_s\, \tau\left( \frac{m}{m_s} \right) \right\rfloor,
\end{equation}
where $m_s=\frac{\nu_s}{\nu_\tau}$ is the number of sampling steps per delay period, $\nu_s$ is the sampling frequency of the FPGA, and $\tau(t)$ is the time-varying delay, where $t$ is measured in delay periods, and thus we have $\tau(t+1)=\tau(t)$.
For our system we consider a sinusoidal time-varying delay
\begin{equation}
\tau(t) = \tau_0 + \frac{A}{2\pi} \sin(2\pi\,t),
\end{equation}
where $\tau_0$ is the mean delay and $A$ is the delay amplitude, which are both measured in delay periods.
The trajectory $y(t)$ in continuous time $t$, where $t$ is measured in delay periods, is connected to the trajectory $y[m]$ in discrete time by
\begin{equation}
y(t+t_0)= y\left[ \lfloor t\,m_s \rfloor \right],
\end{equation}
where $t_0$ is the initial time of the continuous trajectory $y(t)$, such that $y(t_0)=y[0]$.
Consequently, our system in principal has to be modeled by a system with discrete time, since the low pass filter and the time-varying delay are realized digitally by the FPGA.
However, since the digital low pass filter is a good approximation for an analog one, our system can be well described by a DDE with piecewise constant retarded argument $R(t)$.

It is known that the discretization of the retarded argument $R(t)$ can increase the complexity of the dynamics \cite{busenberg_survey_1991,carvalho_nonlinear_1988,norris_period_1991,insperger_semidiscretization_2015}.
The digitally implemented time-varying delay can lead to different dynamics as one expects in the case of a continuous delay, since in this case the access map $R(t)$ is spatially discretized.
It is known that the spatial discretization of a dynamical system may drastically changes the dynamics of the system, such as chaotic dynamics in the continuous system changes into periodic dynamics in the spatially discretized system \cite{binder_simulating_1986,beck_effects_1987,grebogi_roundoff-induced_1988,bastolla_attraction_1997}.
It is also known for spatially discretized maps that spurious stable fixed points or periodic orbits occur, which are close to the stable and unstable fixed points or periodic points of the continuous map \cite{binder_machine_1991,binder_unstable_2003}.
Since the intensity of the output of the interferometer can be measured only with a finite precision, the optical system itself can also be considered as a spatially discretized system.
However, in our case discretization steps of the intensity are small compared to the intensity variation, such that this effect can be neglected.

For the whole paper we have chosen parameters such that $\nu_s \gg f_{\text{cutoff}}$ and have chosen a sampling rate $\nu_s$ that is commensurate with the delay frequency.
Roughly speaking, the system is unable to follow the discontinuities of the delay, since the characteristic frequency of the system is limited by the low pass filter to a value much lower than the sampling rate.
In detail, Eq.~\eqref{eq:digfilter}, where $f_{\text{cutoff}} / \nu_s \approx 0.032 \ll 1$, can be viewed as an autoregressive moving average of the delayed feedback $I[m]$, such that fast variations of $I(m)$ due to the discontinuities of the delay are smoothed.
Consequently, our system can be modelled well by a DDE with continuously varying delay.

In Section III, we added controlled noise to our opto-electronic oscillator to experimentally study the robustness of Laminar Chaos to different noise levels. At each time step, we add noise to the normalized intensity measured by the ADC, such that Eq.~\eqref{eq:mzmod} becomes
\begin{equation}
\label{eq:mzmodnoise}
I[m]=f( y[m-\tau_s[m]] ) + \zeta\xi_m,
\end{equation}
where $\zeta$ is the strength of the external noise and $\xi_m$ is discrete Gaussian white noise with zero mean and $\langle \xi_m \xi_{m'} \rangle = \delta_{m,m'}$, where $\delta_{m,m'}$ is the Kronecker delta.
The white noise is generated by function \texttt{randn} from the \emph{NumPy} library.

Equations~\eqref{eq:digfilter}, \eqref{eq:digdelay}, and \eqref{eq:mzmodnoise} can also serve as a model of the experimental system with inherent noise with $\zeta=0.001$. We now describe how we obtained this value for the noise strength.

\begin{figure}[!ht]
	\centering
	\includegraphics[width=0.4\textwidth]{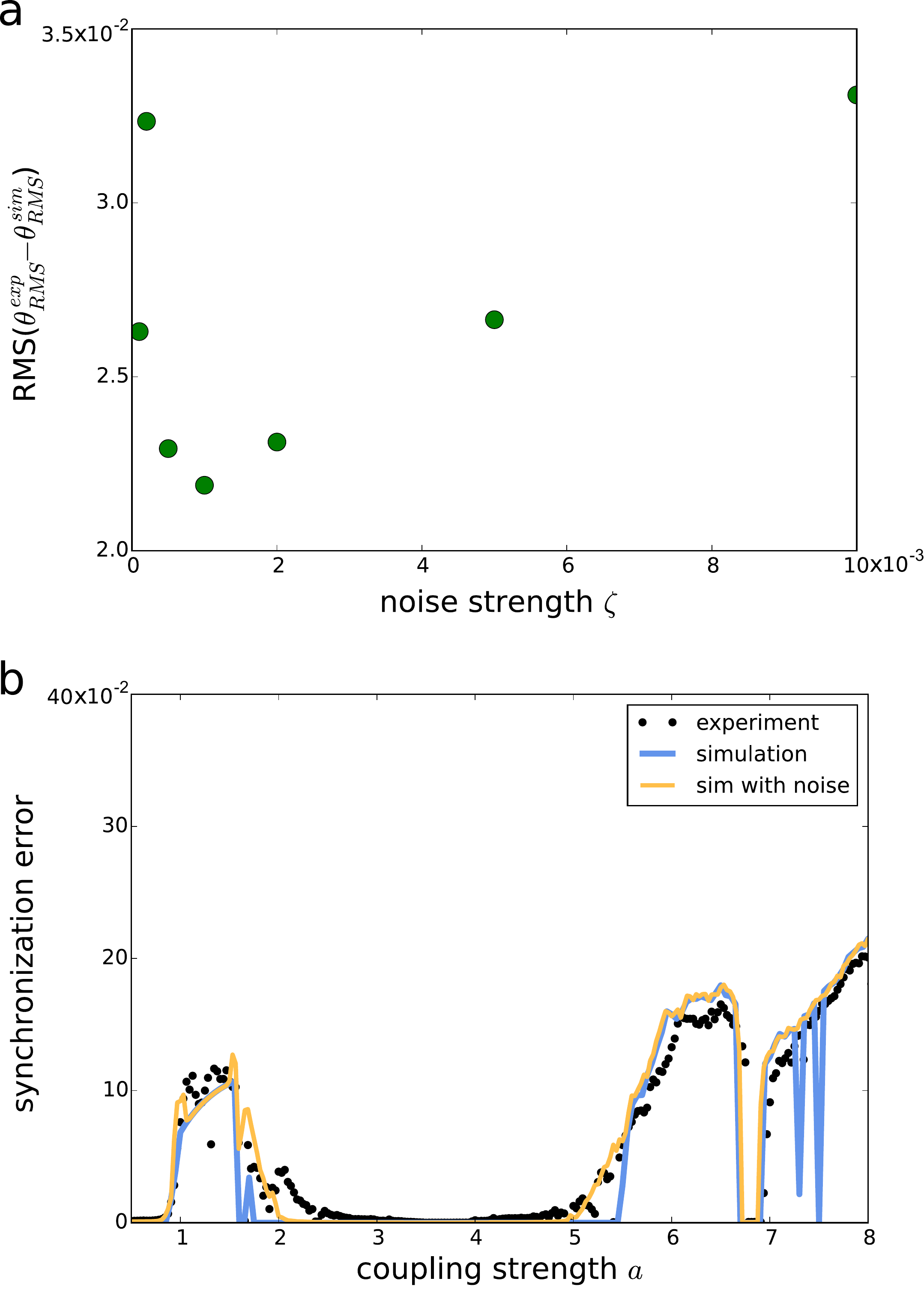}
	\caption{\textit{a} RMS error between the experimentally measured synchronization error and simulated synchronization errors with different noise strengths $\zeta$. \textit{b} Comparison of experimental and simulated synchronization errors without noise and with $\zeta=0.001$.  }
	\label{fig:SyncError}
\end{figure}

We use the exact same apparatus used for our observations of Laminar Chaos; however, we reconfigure the FPGA as described in ~\cite{hart_experiments_2017}, so that the system behaves as two coupled truly identical opto-electronic maps. This system can be modeled as  
\begin{equation}
\begin{aligned}
\label{eq:coupledmaps}
y_0[m+1] =& \beta ( f(y_0[m]) + \zeta\,\xi_m^0 ) + a ( f(y_1[m]) + \zeta\,\xi_m^1 )\\
y_1[m+1] =& \beta ( f(y_1[m]) + \zeta\,\xi_m^1 ) + a ( f(y_0[m]) + \zeta\,\xi_m^0 ),
\end{aligned}
\end{equation}
where $f(y)=\sin^2(y+\phi_0)$ and $\xi_m^i$ is Gaussian white noise with zero mean and $\langle \xi_m^i \xi_{m'}^i \rangle = \delta_{m,m'}$. We note that we have grouped all sources of noise together and model them by applying additive Gaussian noise with standard deviation $\zeta$ to each normalized intensity $I_i$ at each time step.

In order to determine the experimental noise strength, we fix $\beta = 3.5$ and $\delta = \frac{\pi}{4}$, and sweep $a$ from 0 to 8. We also simulate Eq.~\eqref{eq:coupledmaps} with different noise strengths $\zeta$. The RMS synchronization error for the two oscillators is defined as
\begin{equation}
\theta_{RMS}(a;\zeta=\zeta') = \sqrt{\frac{\langle(y_0[n]-y_1[n])^2\rangle}{\langle y_0^2[n]\rangle+\langle y_1^2[n]\rangle}},
\end{equation}
where $\langle\cdot\rangle$ refers to an average over all time and $\theta_{RMS}$ is written as a function of the coupling strength $a$ and a fixed noise strength $\zeta=\zeta'$. A typical plot of the RMS synchronization error as a function of coupling strength is shown in Fig.~\ref{fig:SyncError}b.

We then compute the RMS of the difference between $\theta_{RMS}(a;\zeta=\zeta_{intrinsic})$ from the experiment and $\theta_{RMS}(a;\zeta=\zeta')$ from simulations, where now the RMS averaging is done over all coupling strengths $a\in[0,8]$. The result is shown in Fig.~\ref{fig:SyncError}a and shows a clear minimum at $\zeta=0.001$. The synchronization error from the simulation with $\zeta=0.001$ shows good agreement with the experimentally measured result, as shown in Fig.~\ref{fig:SyncError}b, confirming that we can accurately model the noise in our system. Since this system uses the exact same apparatus as the experiment in which we observe Laminar Chaos, we can apply the same noise model.

\section{Determination of the delay period in experimental time series}
\label{sec:apx_delayperiod}

The estimation of the temporal distribution of the variances given by Eq.~\eqref{eq:tdistvar} is very sensitive to errors of the period of the delay $T_\tau$, especially for averages over a long time series.
Let us consider an experimental trajectory $x(\tilde{t})$ and assume that we have determined the estimate $\hat{T}_\tau$ of the delay period $T_\tau$.
The contribution of the error $|\hat{T}_\tau-T_\tau|$ of $T_\tau$ to the error of $\sigma_d^2$ grows with the length $N$ of the time series in delay periods.
This can be exploited for a precise measurement of the delay period $T_\tau$, where the precision grows with the length of the time series.
It follows directly from Eq.~\eqref{eq:tdistvar} that the error of $\sigma_d^2$ reaches the order of $\sigma_d^2$ if the error of $T_\tau$ reaches or exceeds the order of $\frac{T_\tau}{N}$.
If the error of $T_\tau$ is much larger than $\frac{T_\tau}{N}$, the estimate of $\sigma_d^2$ gets close to the constant variance $\bar{\sigma}_d^2$ of the time series for all $t$.
This can be explained by the fact that the time-varying delay in general leads to a quasiperiodic or periodic frequency modulation of the trajectories, where one of the periods is the delay period $T_\tau$ \cite{muller_resonant_2019}.
The quasiperiodic or periodic frequency modulation leads to a quasiperiodic or periodic variation of the amplitude of the approximated derivatives of the trajectories.
Since the estimated delay period $\hat{T}_\tau$ is almost surely incommensurate to the periods of this amplitude modulation, the variance of the approximated derivatives $\sigma_d^2$ sampled with sampling time $\hat{T}_\tau$ is independent of the initial time $t$ of the time average and equals $\bar{\sigma}_d^2$ if the trajectory length is infinite.

\begin{figure}[!t]
\includegraphics[width=0.4\textwidth]{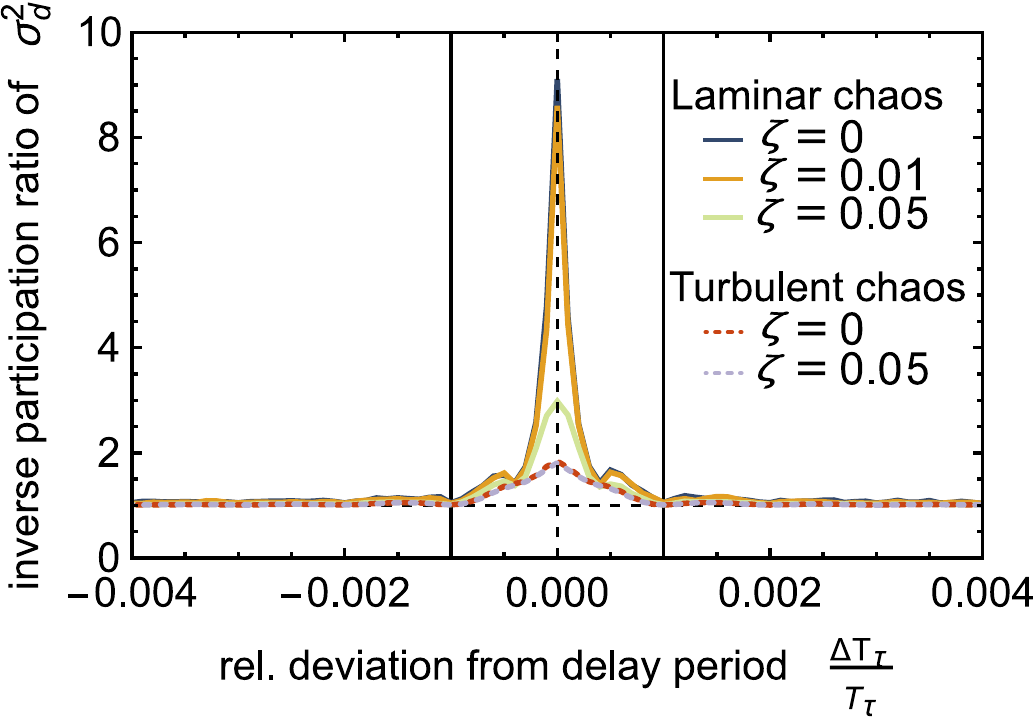}
\caption{\label{fig:delay_period}
Determination of the delay period by computing the inverse participation ratio (IPR) of the temporal distribution of variances $\sigma_d^2$.
Here the IPR of $\sigma_d^2$ in dependence of the relative deviation of the estimated delay period from the exact delay period is shown.
The IPR exhibits a sharp peak at the exact delay period, whereas for larger deviations the IPR is close to one, which indicates a constant temporal distribution of variances $\sigma_d^2$.
The peak width is bounded by the inverse of the length of the time series (in delay periods) $N^{-1}$, which is indicated by the solid vertical lines that are located at $\pm N^{-1}$, where $N=1000$.
}
\end{figure}

If the correct sampling time $T_\tau$ is chosen for the time-average, $\sigma_d^2$ is localized on the time line for Laminar and Turbulent Chaos as shown in Fig.~\ref{fig:tempvard} and Fig.~\ref{fig:tempvardturb}, respectively.
For Laminar Chaos, $\sigma_d^2$ is localized at the transitions between the laminar phases and, for Turbulent Chaos, $\sigma_d^2$ is localized at the high frequency regions that appear with period $T_\tau$.
Thus, the following method for the precise determination of the delay period can be provided.
We consider the following quantity of $\sigma_d^2$, which is known as inverse participation ratio (IPR) in the theory of localization \cite{thouless_electrons_1974}.
\begin{equation}
\label{eq:ipr}
IPR[\sigma_d^2[z]]=\frac{ \int_{0}^1 |\sigma_d^2[z](t)|^4 \, dt }{ \left( \int_{0}^1 |\sigma_d^2[z](t)|^2 \, dt \right)^2 },
\end{equation}
and equals one for constant $\sigma_d^2$ and infinity, if $\sigma_d^2$ consists of delta functions.
To validate the accuracy of $\hat{T}_\tau$, we compute the IPR of $\sigma_d^2$ with $z(t)=x(t\, \hat{T}_\tau)$.
The inverse participation ratio of $\sigma_d^2$ depending on $\hat{T}_\tau$ for the experimental trajectories in Sec.~\ref{sec:exprealize} is visualized in Fig.~\ref{fig:delay_period}.
It exhibits a sharp peak at the delay period $T_\tau$.
Since the error of $T_\tau$ becomes relevant if it is of the order of $\frac{T_\tau}{N}$, the error of the delay period that is determined by this method decreases with increasing $N$.
In other words, the width of the peak of the IPR decreases with increasing $N$, which leads to increasing precision in the estimation of $T_\tau$.

\section{Reconstruction of the nonlinearity fails for Turbulent Chaos}
\label{sec:apprecturb}

In Sec.~\ref{sec:lamchaosnoise} we have demonstrated that Laminar Chaos can be detected by searching for its robust features.
It is characterized by a sequence of laminar phases with periodic alternating duration, which lead to local minima in the temporal distribution of the variance of the approximated derivatives of the trajectories $\sigma_d^2$, which is defined by Eq.~\eqref{eq:tdistvar}.
These laminar phases are connected by a one-dimensional map.
The latter is a key feature of Laminar Chaos, which must be detected for a reliable decision, whether a trajectory shows Laminar Chaos, or not.

\begin{figure}[t]
\includegraphics[width=0.4\textwidth]{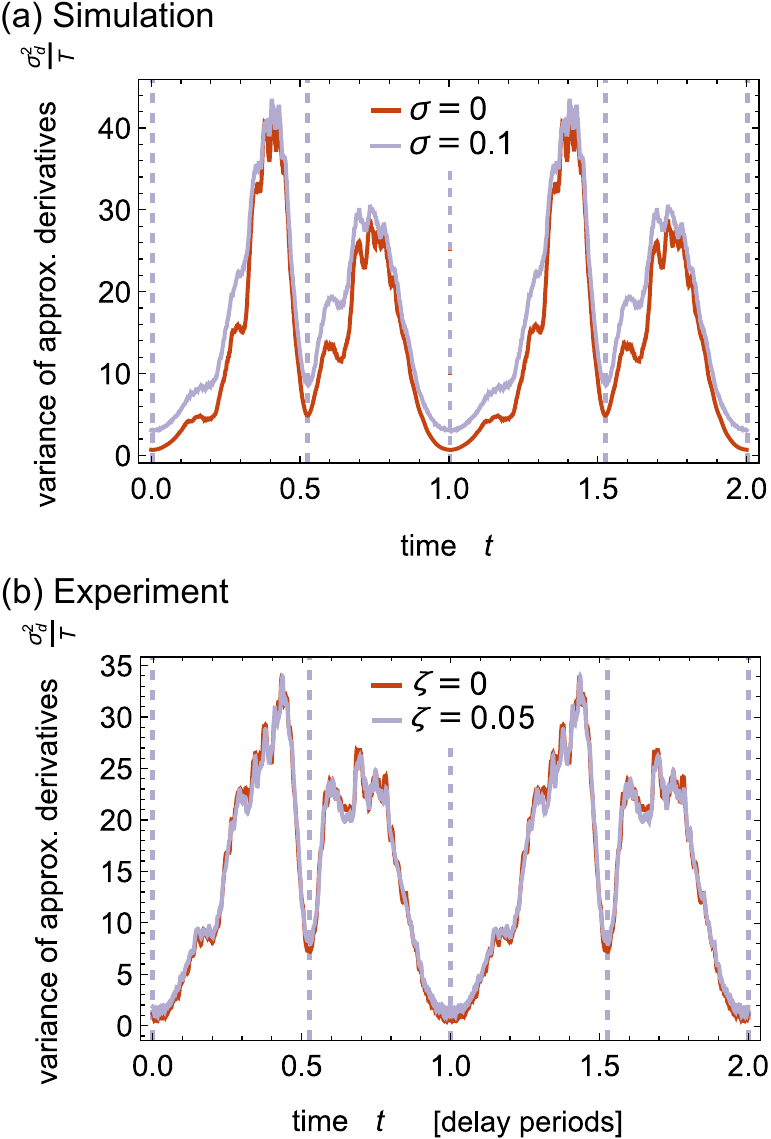}
\caption{\label{fig:tempvardturb}
Temporal distribution of the variance $\sigma_d^2$ (in units of $T$) of the approximated derivatives ($h \approx 0.0034$) of
(a)
laminar chaotic trajectories of Eq.~\eqref{eq:dimlessDDE} with the same parameters as in Fig.~\ref{fig:traj} and
(b)
experimental trajectories for different strengths $\zeta$ of the external noise.
$\sigma_d^2$ is defined by Eq.~\eqref{eq:tdistvar} and is a measure for the fluctuation strength of the trajectory at a specific time relative to the internal clock induced by the time-varying delay.
The temporal distribution of the variance $\sigma_d^2$ in (b) for the experimental time series was shifted, such that one of the numerically computed local minima is located at $t=0$.
The dashed lines in (a) and (b) indicate the local minima, which where determined numerically after denoising $\sigma_d^2$ by applying a Gaussian filter with a standard deviation of $50\,h$.
}
\end{figure}

To emphasize this fact, we demonstrate that Turbulent Chaos may also leads to local minima of $\sigma_d^2$.
These local minima become clearly visible in the case, where the delay is conservative but close to a dissipative delay with a low denominator $q$ of the rotation number $\rho=-\frac{p}{q}$.
Here ``close'' means that the delay parameters are close to parameters that correspond to such a dissipative delay.
As demonstrated in Ref.~\cite{muller_resonant_2019}, a variable delay leads to a frequency modulation of the trajectory similar to the known Doppler effect.
In the case of variable delay systems such as Eq.~\eqref{eq:sysdef} and \eqref{eq:ndimsysdef1}, the signal that is modulated by the Doppler effect is fed back into the system.
For dissipative delay this feedback is resonant, which means that low and high frequency phases appear periodically after $q$ round trips inside the feedback loop due to the mode-locking behavior of the reduced access map, where $q$ is the denominator of the associated rotation number.
The low (high) frequency regions are located in the neighborhood of the attractive (repulsive) periodic points of the access map.
For a conservative delay that is close to a dissipative delay with the rotation number $\rho=-\frac{p}{q}$, the reduced access map has no $q$-periodic points but there are points that remain close after $q$ iterations, which is similar to type-I intermittency (cf. Ref.~\cite{schuster_deterministic_2005}), where nearly periodic motion is interrupted by chaotic bursts.
Consequently, low frequency regions appear approximately after $q$ round trips inside the feedback loop, which leads to the $q$ minima of $\sigma_d^2$ with large depth.
Nevertheless, the values $z_n$ of the trajectory at the local minima of $\sigma_d^2$, which for Laminar Chaos would be considered as intensity levels of the laminar phases, are not connected by a one-dimensional map.
This means that there is no positive $p'\in N$ such that the points $(z_n,z_{n+p'})$ resemble a line.
In Fig.~\ref{fig:tempvardturb} the temporal distribution of the variance $\sigma_d^2$ (in units of $T$) of the approximated derivatives is visualized for (a) trajectories that were computed numerically from Eq.~\eqref{eq:dimlessDDE} and for (b) experimental trajectories.
In both cases the analysis is done for different noise realizations for a conservative delay that is close to a dissipative delay with $\rho=-\frac{3}{2}$.
Besides the minima with a low depth, two minima with comparably large depth are clearly visible.

\begin{figure}[t]
\includegraphics[width=0.4\textwidth]{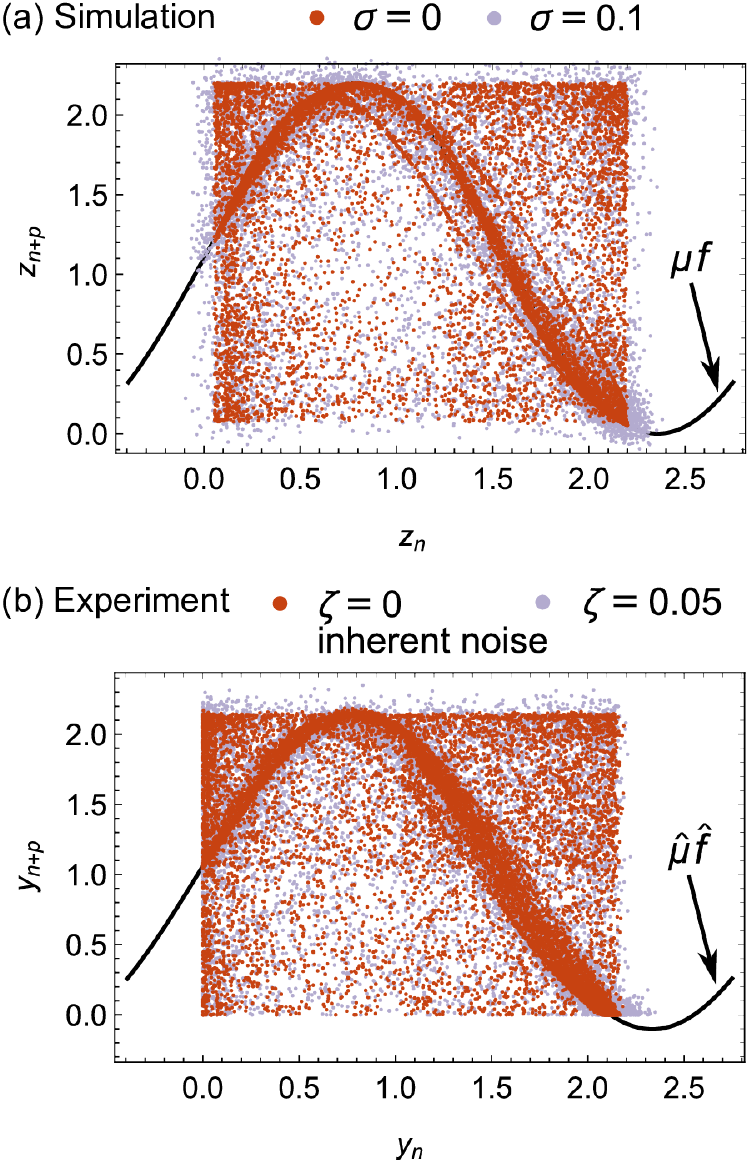}
\caption{\label{fig:nlinrecturb}
Attempt to reconstruct the nonlinearity from trajectories that show Turbulent Chaos for different noise strength, where we considered (a) trajectories of Eq.~\eqref{eq:dimlessDDE} with the same parameters as in Fig.~\ref{fig:traj} and (b) experimental trajectories with the same parameters as in Fig.~\ref{fig:trajexp}.
For numerical (experimental) data, in Fig.~(a) (Fig.~(b)) the value $z_{n+p'}$ ($y_{n+p'}$) is plotted against $z_n$ ($y_n$), where in this case $p'=3$ and the $z_n=z(t^\dagger_n)$ are the values of the trajectories at the periodic sequence of local minima $t^\dagger_n$ of the temporal distribution of the variance $\sigma_d^2$, where only the $q=2$ lowest local minima inside one delay period are considered.
The points $(z_{n},z_{n+p'})$ and $(y_{n},y_{n+p'})$, respectively, do not resemble a line for all positive $p'\in\mathbb{N}$, which leads to a clear distinction from Laminar Chaos in contrast to the temporal distribution of the variance $\sigma_d^2$.
Nevertheless, for $p'=3$ signatures of the nonlinearities $\mu\,f$ and $\hat{\mu}\,\hat{f}$ (black solid lines) are also visible in (a) and (b), respectively, where $\hat{\mu}\,\hat{f}$ is given by the fit to the experimental data in Fig.~\ref{fig:nlinrec}.
This is due to the fact that in our example the conservative delay is close to a dissipative delay, where the access map has the rotation number $\rho=-\frac{p}{q}=-\frac{3}{2}$  (see text).
For this example, an infinite number of points $(z_{n},z_{n+p'})$ or $(y_{n},y_{n+p'})$ would fill a square-shaped set.
}
\end{figure}

In the following, we consider the periodically continued sequence of the locations of the $q$ minima with large depth, which are indicated by the vertical dashed lines in Fig.~\ref{fig:tempvardturb}.
We use them as potential positions $t^\dagger_n$ of the laminar phases and try to reconstruct the nonlinearity from the numerical and experimental time series following the procedure in Sec.~\ref{sec:nlinrec}.
In Fig.~\ref{fig:nlinrecturb} the attempt to reconstruct the nonlinearity for numerical and experimental realizations with different noise strengths is visualized.
Since the points $(z_{n},z_{n+p'})$ (or $(y_{n},y_{n+p'})$, for the experimental data), where the $z_n=z(t^\dagger_n)$ are the values of the trajectory at the potential positions $t^\dagger_n$ of the laminar phases, do not resemble a line, the dynamics can be clearly distinguished from Laminar Chaos.
However, if $p'$ is chosen equal to the numerator $p$ of the rotation number $\rho=-\frac{p}{q}$ of the close dissipative delay, signatures of the nonlinearities appear in the sense that the density of the points $(z_{n},z_{n+p'})$ is large in the vicinity of the graph of the nonlinearity $\mu f(z)$.
This can be explained as follows.

Due to the non-resonant Doppler effect (cf. Ref.~\cite{muller_resonant_2019}), the quasiperiodic dynamics of the access map leads to a quasiperiodic variation of the characteristic time that is associated to the chaotic fluctuations of $z(t)$.
The characteristic time passes through both of the following cases for certain values of $T$ as explained below.
If the characteristic time reaches or exceeds the order of $\Delta t^\dagger_n := |t^\dagger_n-R(t^\dagger_{n+p})|$ in the vicinity of $t^\dagger_n$, the dynamics can be described approximately by the limit map, i.e., we have $z(t^\dagger_{n+p}) \approx \mu f(z(t^\dagger_n))$, and thus $(z_n,z_{n+p})$ is close to the graph of the nonlinearity $\mu f(z)$.
If, in contrast, the characteristic time is much lower than $\Delta t^\dagger_n$ in the vicinity of $t^\dagger_n$, then $z(t^\dagger_n)$ and $z(R(t^\dagger_{n+p}))$ are nearly independent, such that, in general, $(z_n,z_{n+p})$ is not close to the graph of the nonlinearity $\mu f(z)$.
For $T\to\infty$, the characteristic time vanishes, since the function values of the solution $z(t)$ become pairwise independent inside the state interval and thus, only the latter case is possible.
Then, in the absence of noise, the points $(z_n,z_{n+p})$ fill a two-dimensional set $\mathcal{S}$, which is the union of Cartesian products of pairs of the chaotic bands of the map that is defined by the nonlinearity $\mu f(z)$.
If $T$ is large but small enough such that the characteristic time can reach the order of $\Delta t^\dagger_n$, both cases are possible.
The points $(z_n,z_{n+p})$ for which the characteristic time reaches the order of $\Delta t^\dagger_n$ accumulate at the graph of the nonlinearity $\mu f(z)$, whereas the remaining points $(z_n,z_{n+p})$ fill a two-dimensional set and roughly resemble the two-dimensional set $\mathcal{S}$ as shown in Fig.~\ref{fig:nlinrecturb}, where the map defined by the nonlinearity $\mu f(z)$ has one chaotic band.
This is contradicting the features of Laminar Chaos.

\bibliography{ddelamchaosmeas}

\end{document}